\documentclass{nature}
\usepackage{graphicx}
\usepackage{cite}
\usepackage{float}
\usepackage{subcaption}
\usepackage{caption}
\usepackage{amsmath,amssymb,amsfonts}
\usepackage{bbding}
\usepackage{xcolor}
\usepackage{bm}
\usepackage{multirow}
\usepackage{multicol}
\usepackage{textcomp}
\usepackage{ragged2e}
\usepackage{url}
\usepackage{comment}
\usepackage[utf8]{inputenc}
\usepackage[T1]{fontenc}
\usepackage{mathtools}
\usepackage{algorithmic}
\usepackage[colorlinks=true,citecolor=blue,urlcolor=green,bookmarks=false,hypertexnames=false]{hyperref}
\makeatletter
\let\saved@includegraphics\includegraphics
\AtBeginDocument{\let\includegraphics\saved@includegraphics}
\renewenvironment{figure}{\@float{figure}}{\end@float}
\makeatother
\title{Random Quantum Neural Networks (RQNN) for Noisy Image Recognition}

\author{Debanjan Konar$^{1,2}$ \Envelope, Erol Gelenbe$^{3}$ \Envelope, Soham Bhandary$^{4}$, Aditya Das Sarma$^{4}$, and Attila Cangi$^{1,2}$ \Envelope}

\begin{document}

\maketitle
\begin{affiliations}
 \item Center for Advanced Systems Understanding (CASUS), D-02826 G\"orlitz, Germany \Envelope d.konar@hzdr.de, \Envelope a.cangi@hzdr.de
 \item Helmholtz-Zentrum Dresden-Rossendorf (HZDR), D-01328 Dresden, Germany \Envelope d.konar@hzdr.de, \Envelope a.cangi@hzdr.de
 \item Institute of Theoretical and Applied Informatics, Polish Academy of Sciences, Poland, and Laboratoire I3S, University Cote d'Azur, 06100 Nice, France, \Envelope seg@iitis.pl
 \item Department of Electronics \& Telecommunication Engineering, Jadavpur University, kolkata-700092, India
\end{affiliations}

\begin{abstract}
 Classical Random Neural Networks (RNNs) have demonstrated effective applications in decision making, signal processing, and image recognition tasks. However, their implementation has been limited to deterministic digital systems that output probability distributions in lieu of stochastic behaviors of random spiking signals. We introduce the novel class of supervised Random Quantum Neural Networks (RQNNs) with a robust training strategy to better exploit the random nature of the spiking RNN. The proposed RQNN employs hybrid classical-quantum algorithms with superposition state and amplitude encoding features, inspired by quantum information theory and the brain's spatial-temporal stochastic spiking property of neuron information encoding. We have extensively validated our proposed RQNN model, relying on hybrid classical-quantum algorithms via the PennyLane Quantum simulator with a limited number of \emph{qubits}. Experiments on the MNIST, FashionMNIST, and KMNIST datasets demonstrate that the proposed RQNN model achieves an average classification accuracy of $94.9\%$. Additionally, the experimental findings illustrate the proposed RQNN's effectiveness and resilience in noisy settings, with enhanced image classification accuracy when compared to the classical counterparts (RNNs), classical Spiking Neural Networks (SNNs), and the classical convolutional neural network (AlexNet). Furthermore, the RQNN can deal with noise, which is useful for various applications, including computer vision in NISQ devices. The PyTorch code\footnote{https://github.com/darthsimpus/RQNN} is made available on GitHub to reproduce the results reported in this manuscript.
\end{abstract}

\section*{Introduction}
\label{intro}

In the last several decades, artificial neural networks have become indispensable for signal processing, computer vision, and pattern recognition. They have achieved remarkable accuracy in visual pattern recognition. Deep neural networks, in their traditional form, are one of the most effective and far-reaching artificial neural networks currently in use\cite{lecun2015, he2016}. The existing deep learning algorithms are often efficient for training deep neural networks. However, there are limitations in deep neural networks' scalability, adaptability, and usability\cite{madan2022, laborieux2021}. Moreover, traditional deep neural networks require massive quantities of data to construct their intricate mappings through a slow training phase that impairs their ability to retrain and adapt to new input\cite{karuna2021}. These problems are exacerbated by Moore's law's coming to an end as traditional computers approach their physical constraints (e.g., high computational complexity and energy consumption) that will stifle performance gains in the next decades. Furthermore, the sharp rise of artificial intelligence applications throughout various research domains comes at the cost of a massive carbon footprint\cite{dhar2020}. As a result, an emerging issue is finding alternative types of neural networks and computing platforms, including neuromorphic\cite{indiveri2006, goltz2021, roy2019, merolla2014} or near-term quantum hardware\cite{grzesiak2020, arute2019} for training them.

Random Neural Networks (RNNs)\cite{gelenbe1989, gelenbe2020, sompo1988} based on a mathematical model and implementations using probabilistic MOS (P-MOS) circuitry provide a promising pathway to tackle this challenge. RNNs mimic information processing in a biophysical neural network, which communicates and processes sparse and asynchronous binary signals in a massively parallel way. As with biological brain networks, inputs in the RNN are stochastic spike trains which are lossless and energy-efficient. RNNs are the preferred method for many machine-learning applications\cite{yin2017, serrano2020, gelenbe2016, gelenbe2002}, because they are incredibly energy efficient, use low power, and enable rapid inference and event-driven data processing. Moreover, the classical and widely employed RNN architectures\cite{gelenbe2002, gelenbe1999} offer small memory requirements based on limiting approximations that can compact thousands of cells into equivalent transfer functions\cite{yin2017}, limiting the training to a small number of parameters compared with classical deep neural network architectures\cite{lecun1998, lecun2015}. Numerous studies have been conducted on RNN applications in the domains of the Internet of Things and smart cities\cite{javed2017, javed_2017}, Cybersecurity\cite{Serrano2021}, video streaming systems optimization\cite{mohamed2002}, and computer vision\cite{feng1996}. RNNs are composed of neurons that receive stochastic spike signals from external sources such as sensory inputs or neurons. These stochastic spike signals arise from separate Poisson processes with positive rates for excitatory spike signals and negative rates for inhibitory spike signals. However, these signals are integer-valued, hence offer an unbounded range of levels of excitation\cite{gelenbe1999}, unlike classical artificial neural networks, which are binary or limited to the region [$0,1$]. Over the last few decades, a variety of classical training methods for RNNs have been proposed, including reinforcement learning, constrained training prior to conversion, spiking variants of back-propagation, relaxation based on a cost function in Hopfield networks, and auto-associative learning for RNNs\cite{koubi1993, mohamed2002, feng1996}. RNN gradient descent learning has been suggested and studied in various applications\cite{gelenbe1993, gelenbe2002, baster2011}. Despite recent progress, a significant disadvantage of classical RNNs lies in a fact that their accuracy on standard pattern recognition benchmarks falls short of that of machine learning counterparts\cite{lecun2015, feng1996, khera2018} due to the stochastic nature of the random spiking neurons. Another roadblock is a dearth of training algorithms that take advantage of the inherent characteristics of random spiking neurons, which include time-efficient codes and the ability to deal with noisy input and output data.

We tackle these challenges by employing concepts from quantum information science (QIS), which comes with the promise of an exponential advantage for a variety of computational tasks. Recently, quantum computing (QC) has been leveraged for machine learning with the hope that the uncertainty in QC can be a great advantage for probability-based modelling in machine learning, inspiring new research for Noisy Intermediate-Scale Quantum (NISQ) devices. Quantum supremacy has shown that some specific tasks that are otherwise extremely lengthy or even unattainable on classical computers can be quickly solved by quantum computing in just a few seconds\cite{arute2019}, saving enormous amounts in energy consumption, a unique “green” feature compared to classical supercomputers. In comparison to high-performance computing, quantum neural networks are still in their infancy owing to a lack of sufficient qubits for implementation on quantum hardware platforms and noise processes that restrict the number of quantum neural networks. Variational quantum circuits (VQCs)\cite{cerezo2021} have emerged as one of the most effective approaches to deep quantum learning in recent years when used with NISQ devices. With the advent of available quantum computing devices and quantum variational algorithms\cite{cerezo2021}, quantum machine learning research has begun to focus on hybrid classical-quantum algorithms that can be performed in the near-term Noisy Intermediate-Scale Quantum (NISQ) devices. Due to the limited number of qubits, these are paving the way for the eventual practical use of NISQ devices for machine learning applications. In fact, classical neural network models incorporating quantum circuits as sub-routines preclude quantum information processing\cite{gyon2019}. VQCs may be trained using classical optimization approaches, and they are believed to offer certain expressibility advantages over typical neural network topologies\cite{abbas2021, benedetti2019}. Examples include quantum state-based Hopfield networks\cite{reben2018} and their classical implementations using quantum-accelerated matrix inversion, Recurrent Quantum Neural Networks\cite{behera2006}, Quantum Convolutional Neural Networks\cite{cong2019}, and Spiking Quantum Neural Networks\cite{kristen2021, suna2020}.

However, concrete deep quantum neural networks for widespread real-world applications like pattern recognition tasks have not been studied yet due to the limited number of qubits available in quantum simulators and a lack of fault-tolerant systems\cite{aggarwal2018, souza2011}. Hybrid classical-quantum algorithms using VQC\cite{liang2021, chen2021, liu2021, mari2020} can be ideally implemented and trained on classical hardware. We employ this workflow to demonstrate the potential quantum advantages that can be achieved on NISQ devices, even with a limited number of qubits.

In this paper we present a novel classical-quantum framework for RNNs. These random quantum neural networks (RQNNs) turn a limitation of NISQ devices into an advantage. They leverage the inherent noise in NISQ devices due to the limited number of qubits available to encode high-dimensional classical data. The significant contributions of the presented work are summarized below. 
\begin{itemize}
	\item We demonstrate the feasibility of RQNN using a VQC in the dressed quantum layer, a hybrid classical-quantum circuit with gate parameters optimized during training, demonstrating a significant difference from existing classical random neural network models\cite{gelenbe2016, gelenbe2020} in terms of training with random quantum spiking neurons for near-term quantum devices.
	\item Furthermore, the dressed quantum layer in the proposed RQNN model as a VQC can be trained with many parameters in the architecture, resulting in the benefits of flattening local minima, as demonstrated by the proposed RQNN model's convergence.
    \item Using the PennyLane quantum simulator, the proposed RQNN model is tested on the MNIST, FashionMNIST, and KMNIST datasets under noisy conditions. It outperforms classical RNNs\cite{gelenbe2016}, classical Spiking Neural Networks (SNNs)\cite{khera2018}, and the classical CNN model (AlexNet\cite{alexnet2012}). Thereby, we introduce a unique and novel approach for solving daunting challenges in computer vision and pattern recognition.
\end{itemize}
The remaining sections of the manuscript are organized as follows: In the \emph{Results} section, we provide details on the data set, experimental setting, and experimental outcomes. In the \emph{Discussion} section, we assess the results achieved with our proposed RQNN framework and compare them with those of classical neural networks. Based on the superior accuracy achieved with RQNNs, we provide insight into the future paths of quantum machine learning research for NISQ devices. In section \emph{Methods}, we provide technical details of our RQNN framework. To that end, we explain key concepts implemented such as our hybrid classical-quantum algorithms, RNNs, and VQC. Finally, the convergence of RQNNs is demonstrated in the \emph{Appendix}.

\section*{Results}
\label{results:disscuss}

\subsection{Data Set.}
\label{results:dataset}

The MNIST dataset\cite{lecun1998} has been widely used as a benchmark for various computer vision tasks prevalent in character recognition using classical RNNs\cite{gelenbe2016, gelenbe2020}. The MNIST dataset\cite{lecun1998} comprises $28 \times 28$ gray-scale images of $70000$ fashion objects organized into ten categories, each with $7000$ images. The training set contains $60,000$ images, whereas the test set contains $100,00$ images. FashionMNIST\cite{xiao2017} is a dataset\footnote{https://github.com/zalandoresearch/FashionMNIST} which uses the same image size, data format, and training and testing split structure as the original MNIST dataset\cite{lecun1998}. Kuzushi-49\footnote{https://github.com/rois-codh/kmnist} is a dataset produced from the KMNIST dataset\cite{clanuwat2018}, which can be used instead of MNIST, the most well-known dataset in the machine learning community. We have decreased the size of the datasets as mentioned above to four classes for training in our experiments due to the limited number of qubits offered by the PennyLane Quantum simulator. Images are corrupted with salt and pepper noise with probability $0.02, 0.04$, to $0.18$, and $0.2, 0.4$, to $1.0$, Gaussian noise with standard deviation $0.01, 0.02$, to $0.09$, and $0.1, 0.2$, to $0.9$, Rayleigh noise with scale $0.01, 0.02$, to $0.09$, and $0.1, 0.2$, to $0.9$, uniform noise with low = $0$ and high = $0.01, 0.02$, to $0.09$, and $0.1, 0.2$, to $.9$, and Perlin noise of resolution $1, 7$, and $14$. The Sneaker, AnkleBoot, Bag, and Shirt classes are used for training in the FashionMNIST\cite{xiao2017} dataset, whereas only $6$, $7$, $8$, and $9$ digits are considered in the MNIST dataset\cite{lecun1998}.
The datasets used for noisy image classification are listed in Table~\ref{tab1}: MNIST\cite{lecun1998}, FashionMNIST\cite{xiao2017} and KMNIST\cite{clanuwat2018}.
\begin{table}
\footnotesize
	\begin{center}
	\caption{Details of the MNIST\cite{lecun1998}, FashionMNIST\cite{xiao2017} and KMNIST\cite{clanuwat2018} datasets (Training-Validation and Test dataset) used for image classification.}
		\begin{tabular}{p{40pt}p{1pt}p{60pt}p{1pt}p{50pt}}
			\hline	
			\multicolumn{1}{p{40pt}}{\centering{\textbf{Dataset}}} &
			\multicolumn{1}{p{1pt}}{}&
			\multicolumn{1}{p{60pt}}{\centering{\textbf{Training and Validation}}} &
			\multicolumn{1}{p{1pt}}{} &
			\multicolumn{1}{p{40pt}}{\textbf{Test Dataset}} \\
			\hline
            MNIST & & $954$ & &	$7073$ \\
            FashionMNIST & & $954$ & & $7172$ \\
            KMNIST & & $954$ & & $5512$\\
            \hline
			\label{tab1}
		\end{tabular}
	\end{center}
\end{table}

\subsection{Experimental Settings.}
\label{experiment:result}

\begin{figure}[htbp]
\centering
 \subcaptionbox{MNIST}{\includegraphics[scale=0.35]{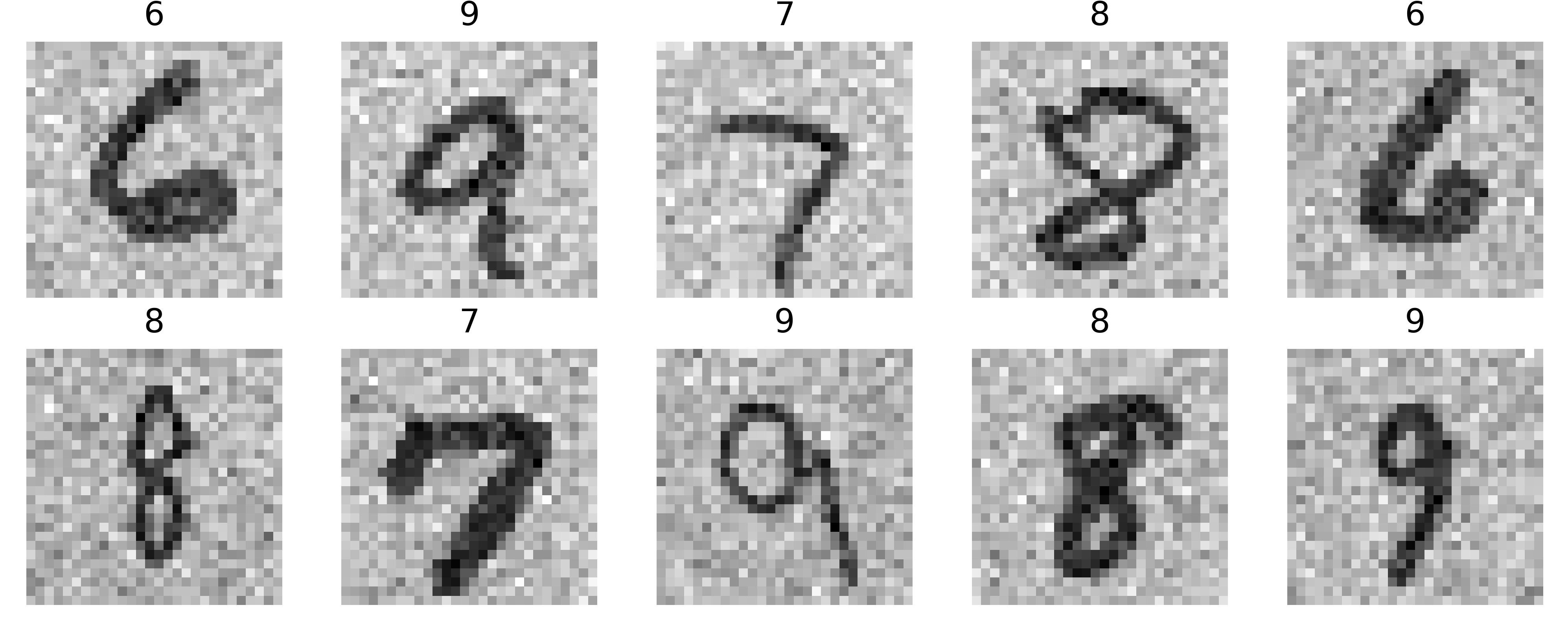}}
 \subcaptionbox{FashionMNIST}{\includegraphics[scale=0.35]{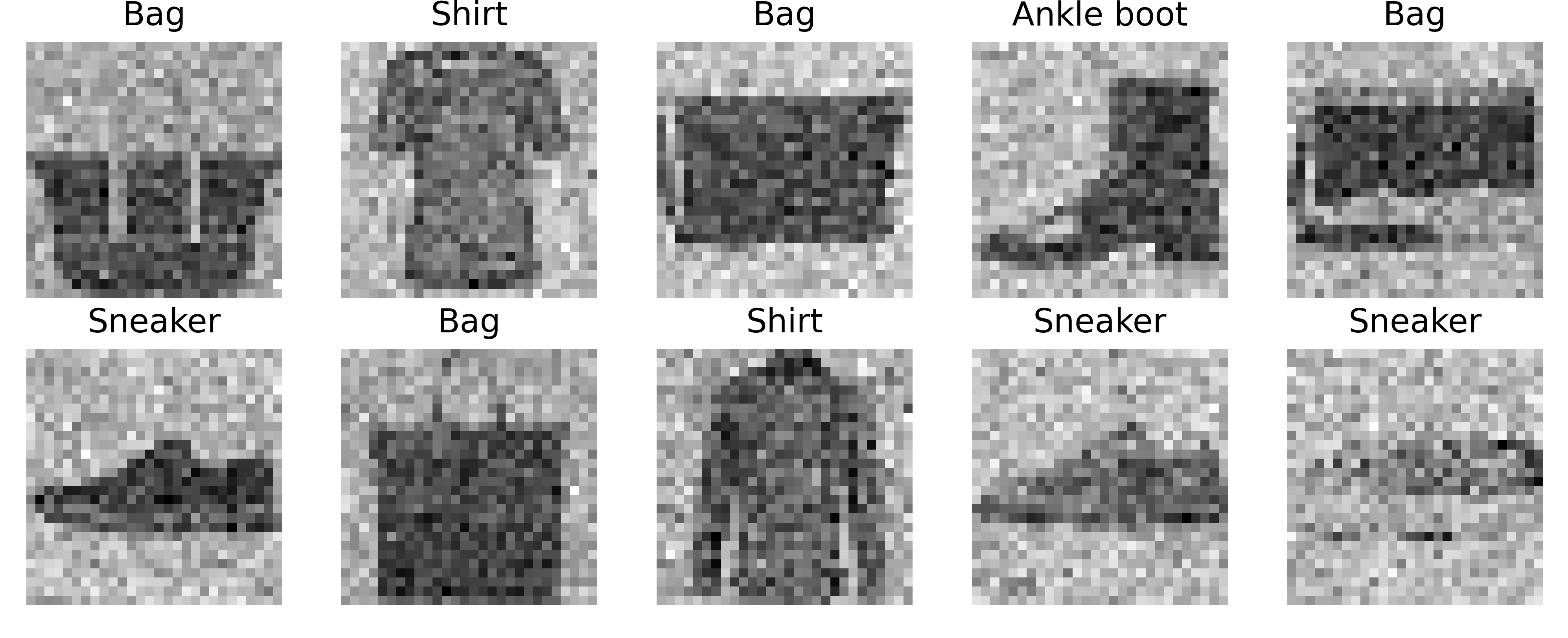}}
  \subcaptionbox{KMNIST}{\includegraphics[scale=0.35]{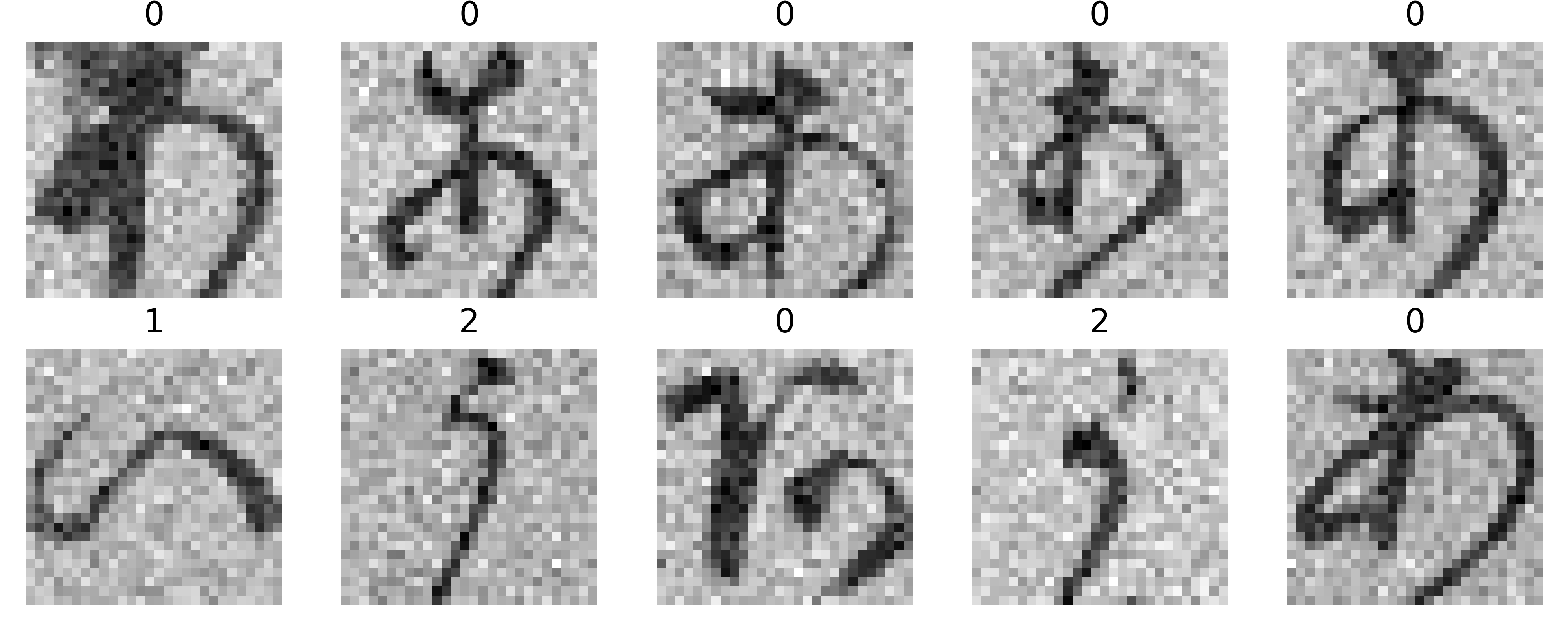}}
	\caption{Gaussian noise with standard deviation of $0.3$ imposed on the images from the (a) MNIST\cite{lecun1998}, (b) FashionMNIST\cite{xiao2017} and (c) KMNIST\cite{clanuwat2018} datasets.}
	\label{fig:noisyImages}
\end{figure}

Numerical experiments have been conducted using our proposed RQNN model, classical RNN\cite{gelenbe2016}, classical SNN\cite{khera2018}, and AlexNet\cite{alexnet2012} to classify gray-scale images in noisy situations. Due to the limited number of qubits available in the PennyLane Quantum simulator \footnote{https://pennylane.ai/}, the size of the MNIST, FashionMNIST, and KMNIST datasets has been decreased to four classes for training. Hence, we consider all the models (proposed RQNN model, classical RNN\cite{gelenbe2016}, classical SNN\cite{khera2018} and AlexNet\cite{alexnet2012}) in four-class classification problems due to the limited number of qubits available for RQNN. Hence, it differs visually from the standard classical RNN, SNN and Alex-Net models. However, testing in noisy settings is carried out to demonstrate the suggested model's robustness in dealing with noise using noisy versions of unseen test images. We conducted the experiments on the high performance computing facility of the Helmholtz-Zentrum Dresden-Rossendorf (HZDR)  (Hemera cluster) which comprises an Nvidia Tesla $V100-SXM2$ GPU Cluster with $32$GB RAM and $640$ Tensor cores with eight cores of Intel(R) Xeon(R) CPU E5-2683 v4 @ 2.1GHz. The proposed RQNN, classical RNN\cite{gelenbe2016}, and AlexNet\cite{alexnet2012} have been developed using Pytorch. In Figure~\ref{fig:noisyImages}, we illustrate the sample input (noisy images) with a dimension of $28 \times 28$ from the KMNIST, MNIST, and FashionMNIST datasets.
Seven hundred eighty-four input features ($28 \times 28$) from noisy input images (FashionMNIST, MNIST, and KMNIST) are applied to the linear input layer of the multi-layer classical RNN, RQNN, classical SNN\cite{khera2018} and $128$ output features are generated after temporal pooling over spikes produced by the spiking version of the activation function ReLU. The pre-input layer of the proposed RQNN framework receives $10$ input features obtained after the temporal pooling layer acts on $128$ features from the preceding layer of the multi-layer traditional RNN architecture\cite{gelenbe2016}. The VQC layer employs a qubit count of $n=6$. The fully connected (FC) parametrized quantum layers of the classical RNN\cite{gelenbe2016} and SNN model\cite{khera2018}, AlexNet\cite{alexnet2012}, and the proposed RQNN model are rigorously trained using the \emph{Adam} optimizer with a maximum of $30$ epochs, an initial learning rate of $0.003$, $\alpha 1 = 0.81$ and $\alpha 2 = 0.88$, and a minimum batch size of $32$. The Figure~\ref{fig:Train_accuracy} and Figure~\ref{fig:Train_loss} show how the accuracy and loss of the proposed RQNN model improve when it is trained using $5$-fold cross-validation.

\subsection{Image Classification with RQNN.}
\label{RQNN:class}

\begin{figure*}[htbp]
	\centering
	\includegraphics[scale=0.20]{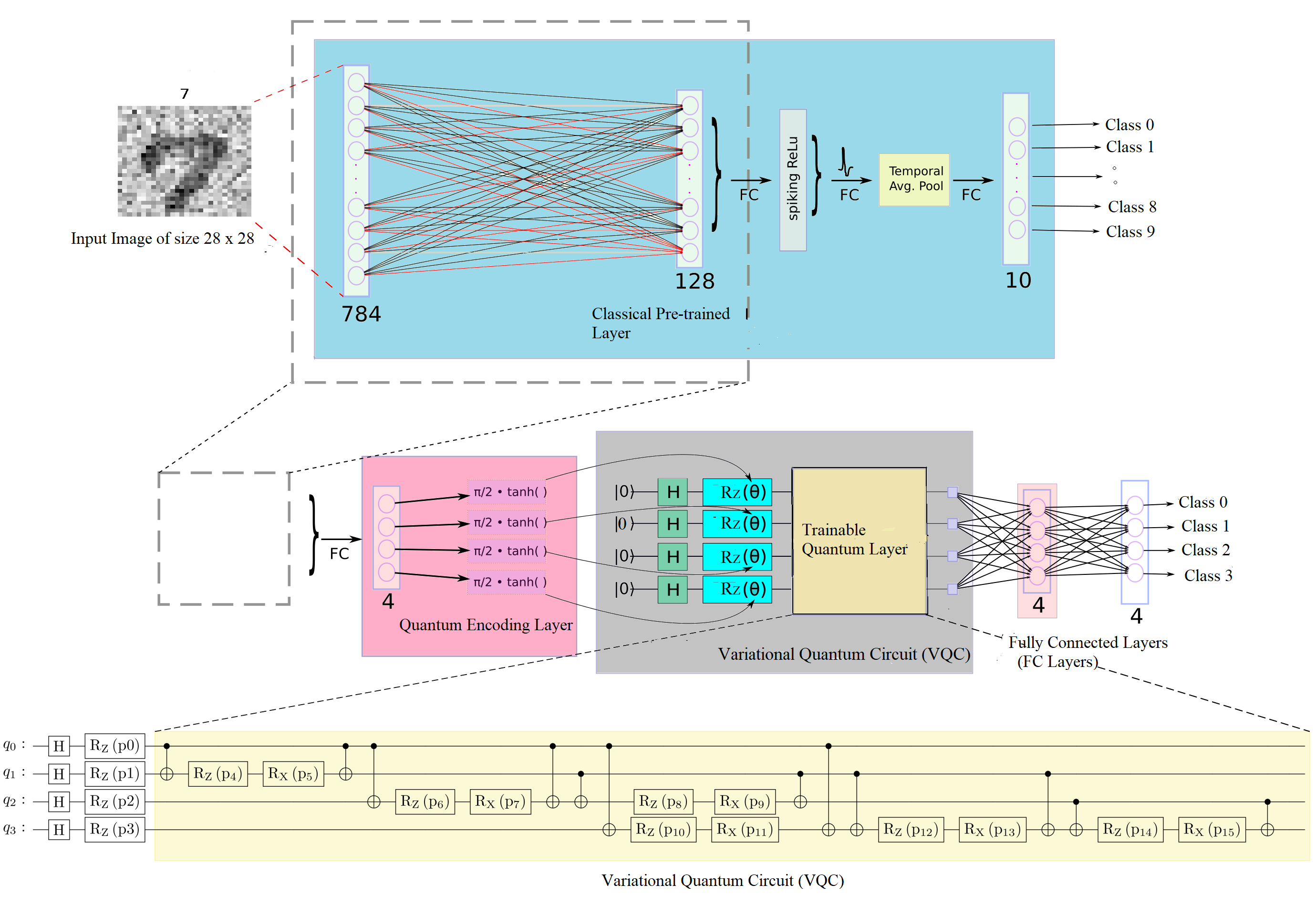}
	\caption{Random Quantum Neural Networks (RQNN) with VQC [The encoding part comprises the Hadamard gate $H$ and single qubit rotation gates $\mathcal{R}_x(\omega)$ and $\mathcal{R}_z(\omega)$ representing rotations along the x-axis and z-axis, respectively and to produce an impartial superposition state, the Hadamard gate $H$ is introduced. The variational portion includes CNOT gates for entangling quantum states from individual qubits, as well as $\mathcal{R}_x(\omega)$ and $\mathcal{R}_z(\omega)$ parameters for the universal single-qubit rotation gate, which must be learned ($p_0, p_1, \cdots p_{15}$). The quantum measurement is conducted on qubits for unitary rotations applied to neighbouring qubits.]}
	\label{fig:RQNN}
\end{figure*}
Recently, quantum computing has been leveraged for machine learning with the hope that the uncertainty in quantum computing can become a great advantage for probability-based modelling in machine learning, inspiring a new line of research at the intersection of quantum computing and neuromorphic computing. In the proposed RQNN architecture, a classical RNN based on a multi-layer perceptron model is used as shown in Figure~\ref{fig:RQNN}, followed by a dressed quantum circuit (VQC) with $m$ qubits. The dressed quantum layers (VQC layers) have a depth of $n$. The RNN's classical layers are based on the random spiked neural networks\cite{gelenbe2016}. With the proposed RQNN model, input gray-scale image features are initially fed into the classical layers of the RNN for image recognition. The random spikes are averaged across time with the temporal pooling layer followed by quantum encoding in terms of the dressed quantum layers. 

\begin{figure}[htbp]
	\centering
 \subcaptionbox{FashionMNIST}{\includegraphics[width=3in]{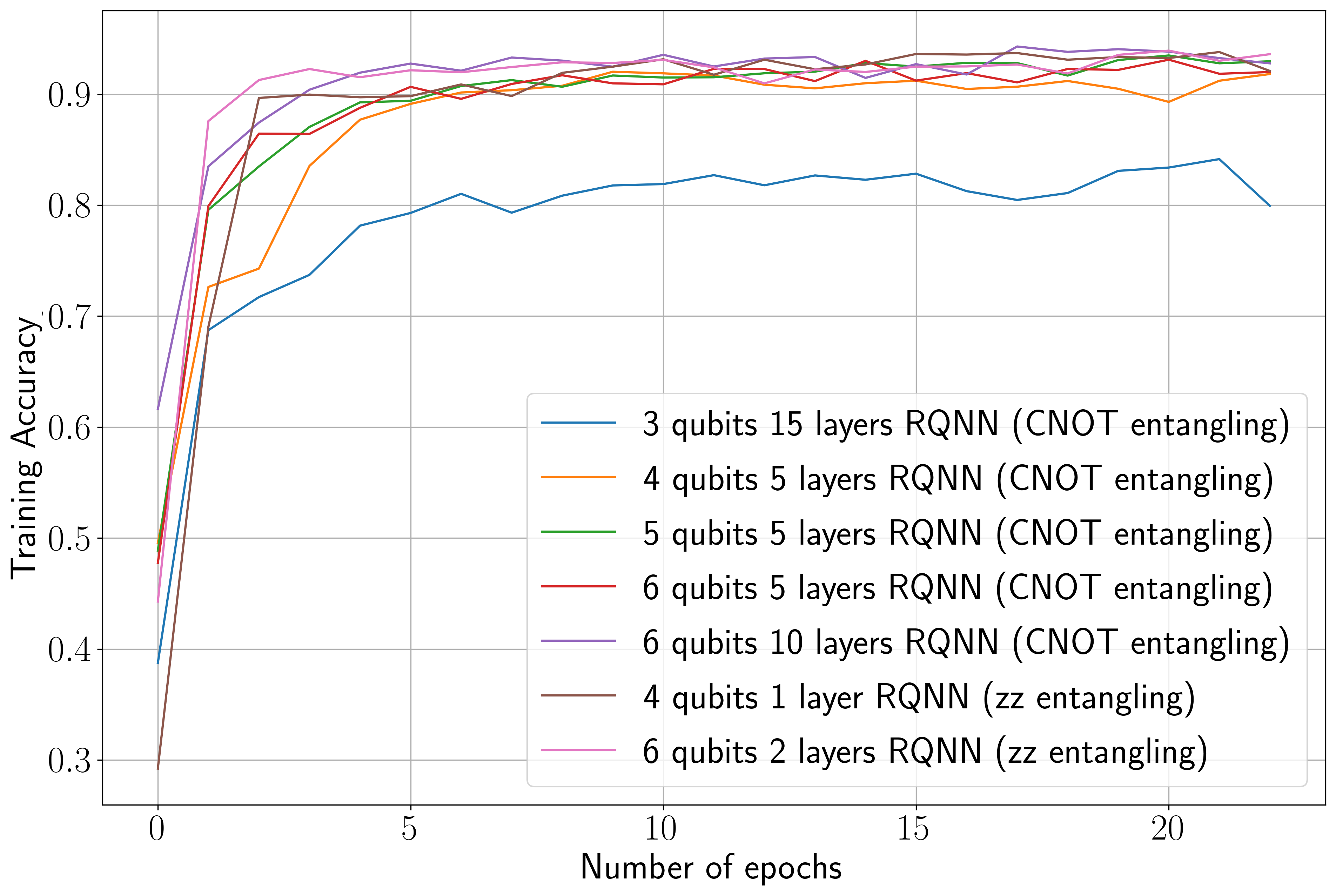}}
 \subcaptionbox{MNIST, FashionMNIST and KMNIST}{\includegraphics[width=3in]{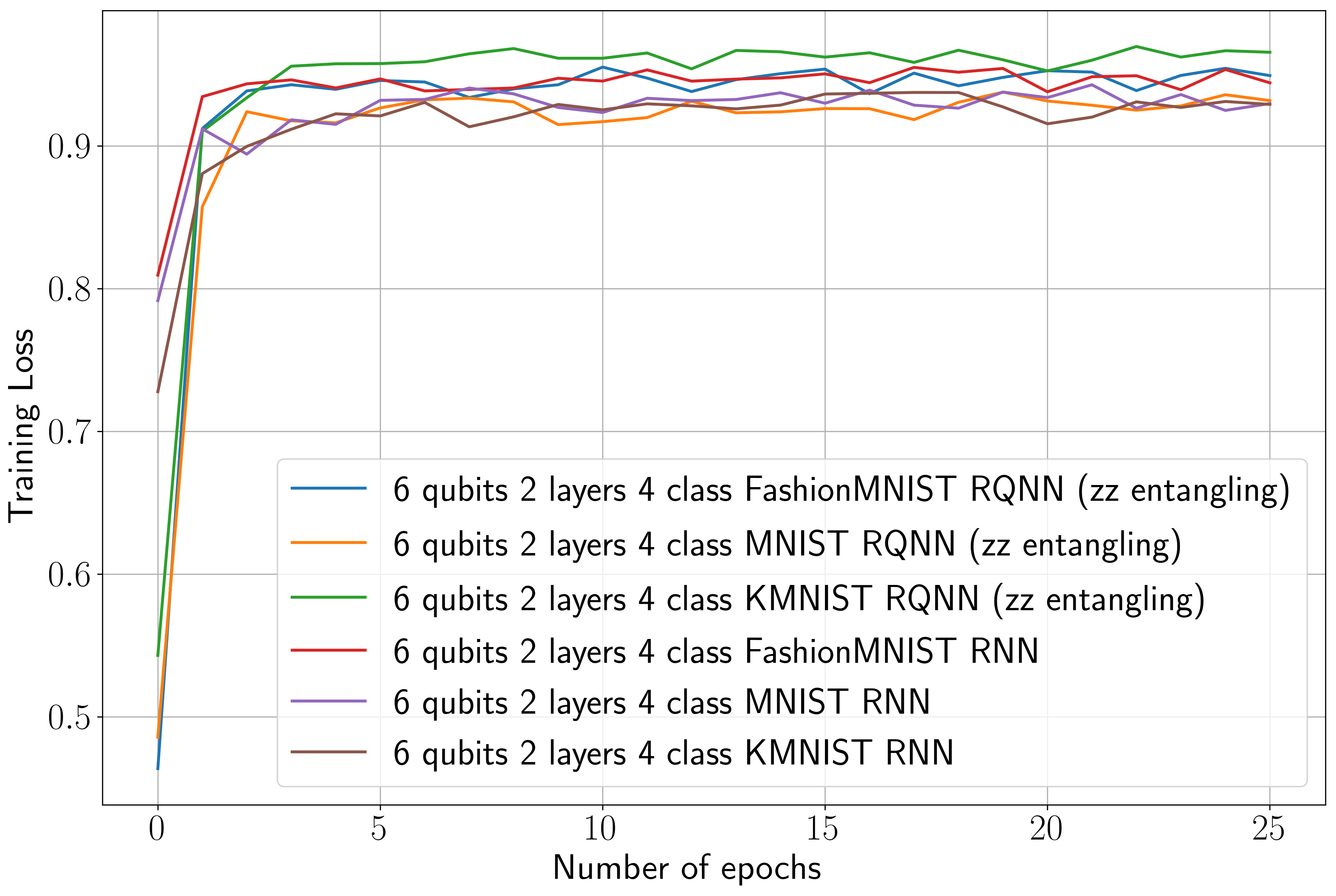}}
 	\caption{The training accuracy of RQNN is simulated on various numbers of qubits and layers of ansatz for (a) $5$ classes of FashionMNIST dataset\cite{xiao2017} (b) $4$ classes of MNIST\cite{lecun1998}, FashionMNIST\cite{xiao2017} and KMNIST\cite{clanuwat2018} datasets.}
 \label{fig:Train_accuracy}
 \end{figure}

\begin{figure}[htp]
	\centering
 \subcaptionbox{FashionMNIST}{\includegraphics[width=3in]{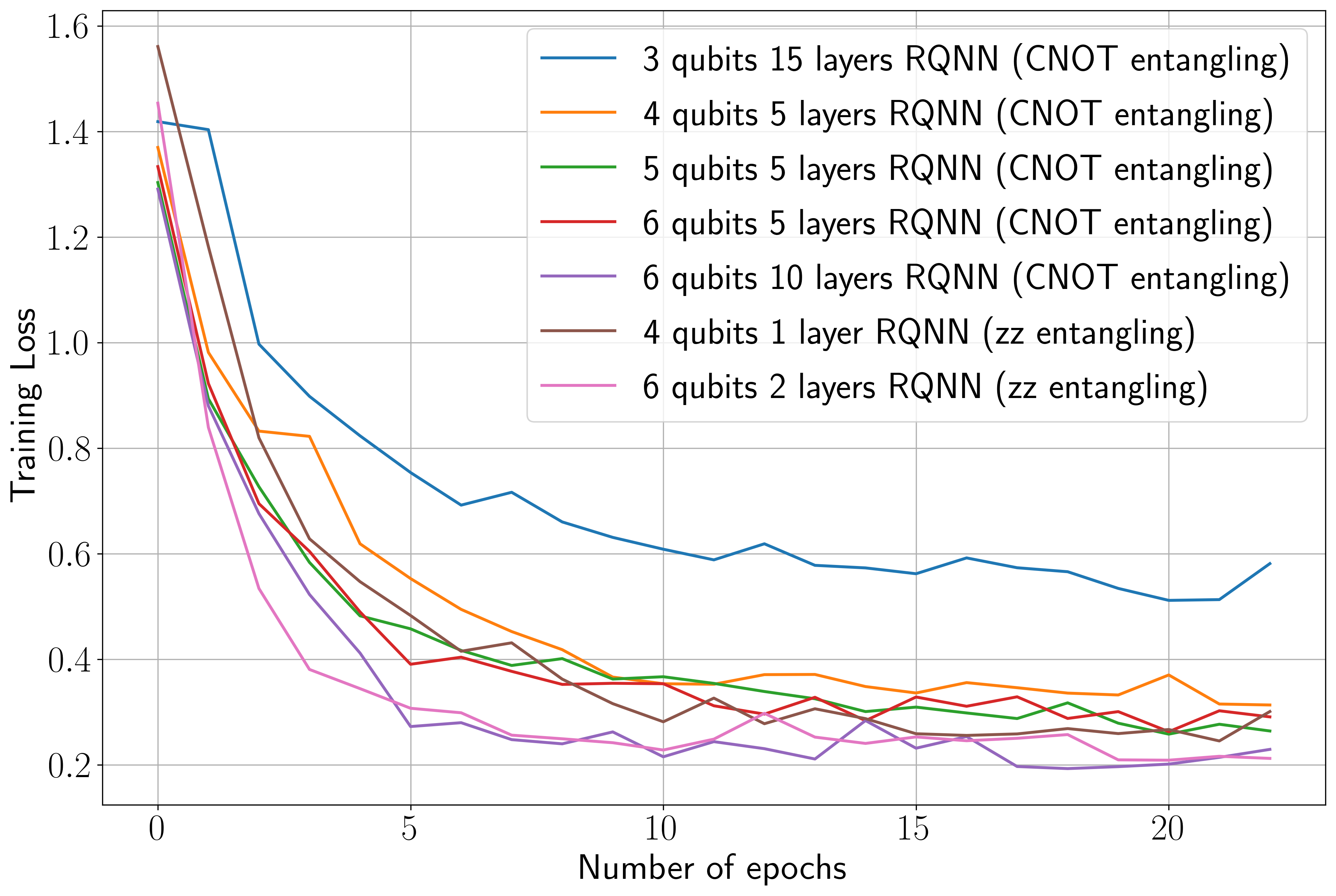}}
 \subcaptionbox{MNIST, FashionMNIST and KMNIST}{\includegraphics[width=3in]{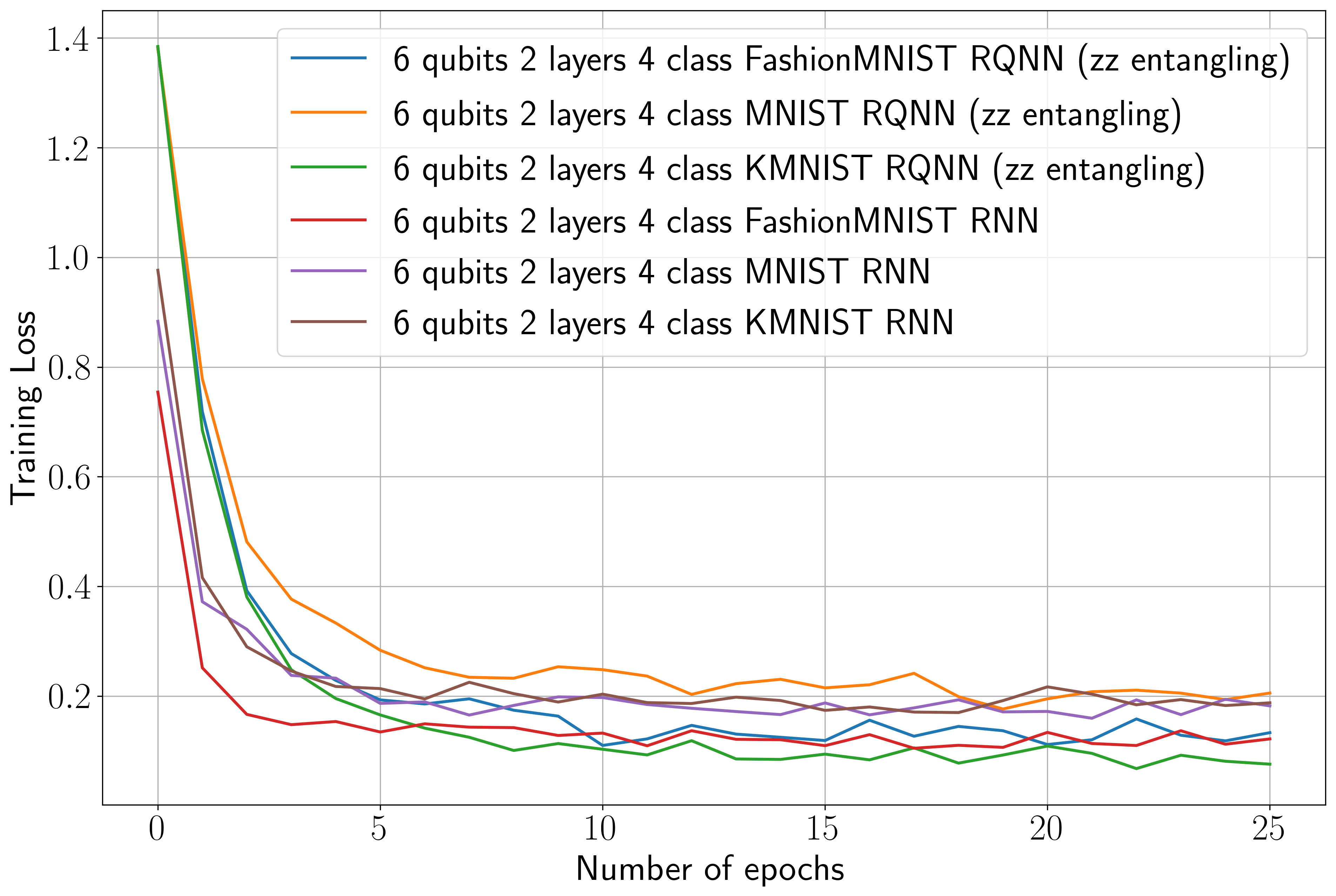}}
 	\caption{The training loss of RQNN is simulated on various numbers of qubits and layers of ansatz for (a) $5$ classes of FashionMNIST dataset\cite{xiao2017} (b) $4$ classes of MNIST\cite{lecun1998}, FashionMNIST\cite{xiao2017} and KMNIST\cite{clanuwat2018} datasets.}
 	\label{fig:Train_loss}
 \end{figure}

Following the four-class classification, the quantum data is encoded into classical bits using a quantum measurement. We note that the proposed RQNN model is considered for four-class classification problems in this work due to the limited number of qubits available. Hence, it differs visually from the standard RQNN model, as shown in Figure~\ref{fig:RQNN}. We employed a hybrid classical-quantum framework to evaluate the proposed RQNN model for noisy image classification on a PennyLane Quantum simulator (real quantum computing hardware). At the classification layers, we employed cross-entropy loss to classify images. The loss function ($\mathcal{L}_{\theta}$) is determined by the RQNN model's hyperparameters $\theta$:
\begin{equation}
   \operatorname*{argmin}_\theta \mathcal{L}_{\theta} = \sum_{j}^{\mathcal{C}} [\gamma_j \log z (\omega_j) + (1-\gamma_j) \log \{1-z (\omega_j)\}]
\end{equation}
which is a proxy for determining the accuracy of the RQNN. For a given input angle $\omega_j$, a fully connected (FC) layer is expected to produce $z(\omega_j)$ with a target outcome of $\gamma_j$ in terms of the set of network hyper-parameters $\theta$. 
The training accuracy and loss function of this RQNN model are shown in Figure~\ref{fig:Train_accuracy} and Figure~\ref{fig:Train_loss}, respectively, using the MNIST\cite{lecun1998}, FashionMNIST\cite{xiao2017} and KMNIST\cite{clanuwat2018} datasets.
 
\subsection{Experimental Results.}
\label{result:exp}

In the current experimental setup, the suggested RQNN, the classical RNN model\cite{gelenbe2016}, the classical SNN\cite{khera2018}, and AlexNet\cite{alexnet2012} are applied, and the numerical results are reported in terms of a statistical analysis. In extensive tests, large sets of the MNIST\cite{lecun1998}, FashionMNIST\cite{xiao2017}, and KMNIST\cite{clanuwat2018} datasets have been used as training, validation, and test sets. The proposed RQNN, the classical RNN model\cite{gelenbe2016}, the classical SNN\cite{khera2018}, and AlexNet\cite{alexnet2012} are trained using images from the MNIST\cite{lecun1998}, FashionMNIST\cite{xiao2017}, and KMNIST\cite{clanuwat2018}. They are, however, evaluated on unseen noisy versions of test images from the datasets to show that the suggested model is more resilient than its classical equivalent and classical neural network-based models. Table~\ref{tab2} summarizes the numerical results obtained on the test datasets affected by salt and pepper noise with a probability $0.3$, Gaussian noise with a standard deviation of $0.3$, Rayleigh noise with a scale of $0.3$, uniform noise with an intensity of $0.3$, and Perlin noise with a resolution of $14 \times 14$, using the suggested RQNN, classical RNN model\cite{gelenbe2016}, classical SNN\cite{khera2018} and AlexNet\cite{alexnet2012} architectures. The presented work employs the mean accuracy ($ACC$), dice similarity score ($DS$), positive prediction value ($PPV$), and sensitivity ($SS$) as evaluation measures. It has been observed from the numerical data reported in Table~\ref{tab2} that the proposed RQNN is superior to the classical RNN model\cite{gelenbe2016}, the classical SNN\cite{khera2018}, and the AlexNet model\cite{alexnet2012} in handling all types of noise except Perlin noise. Furthermore, we have conducted a two-sided paired Wilcoxon signed-rank test\cite{conover1999} with a significance threshold of $\alpha = 0.05$ to show that the proposed RQNN model outperforms the classical RNN model\cite{gelenbe2016}, classical SNN\cite{khera2018}, and AlexNet\cite{alexnet2012}.

\begin{figure}[htbp]
	\centering
 \subcaptionbox{Gaussian noise}{\includegraphics[width=2in]{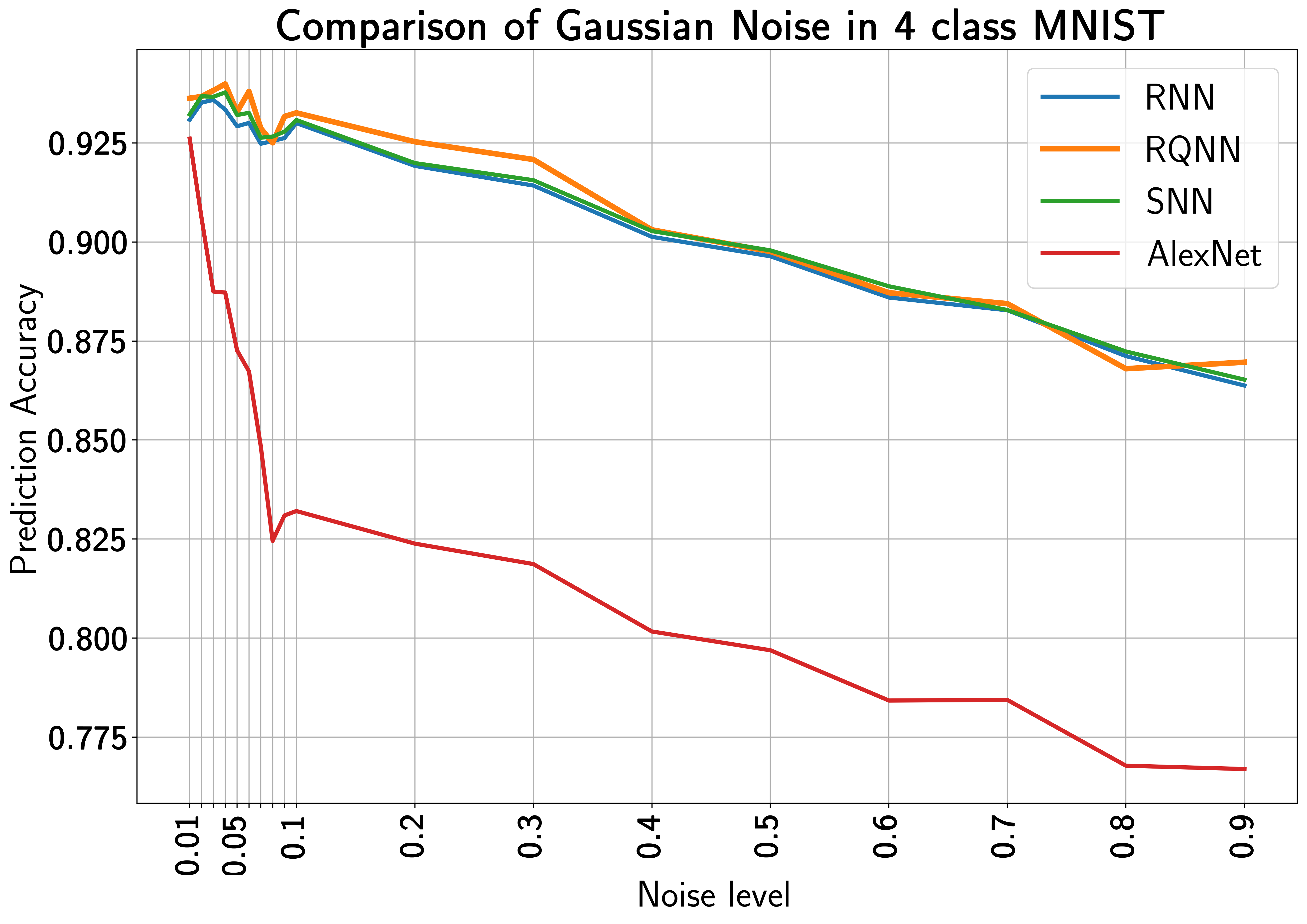}}
 \subcaptionbox{Salt and Pepper noise}{\includegraphics[width=2in]{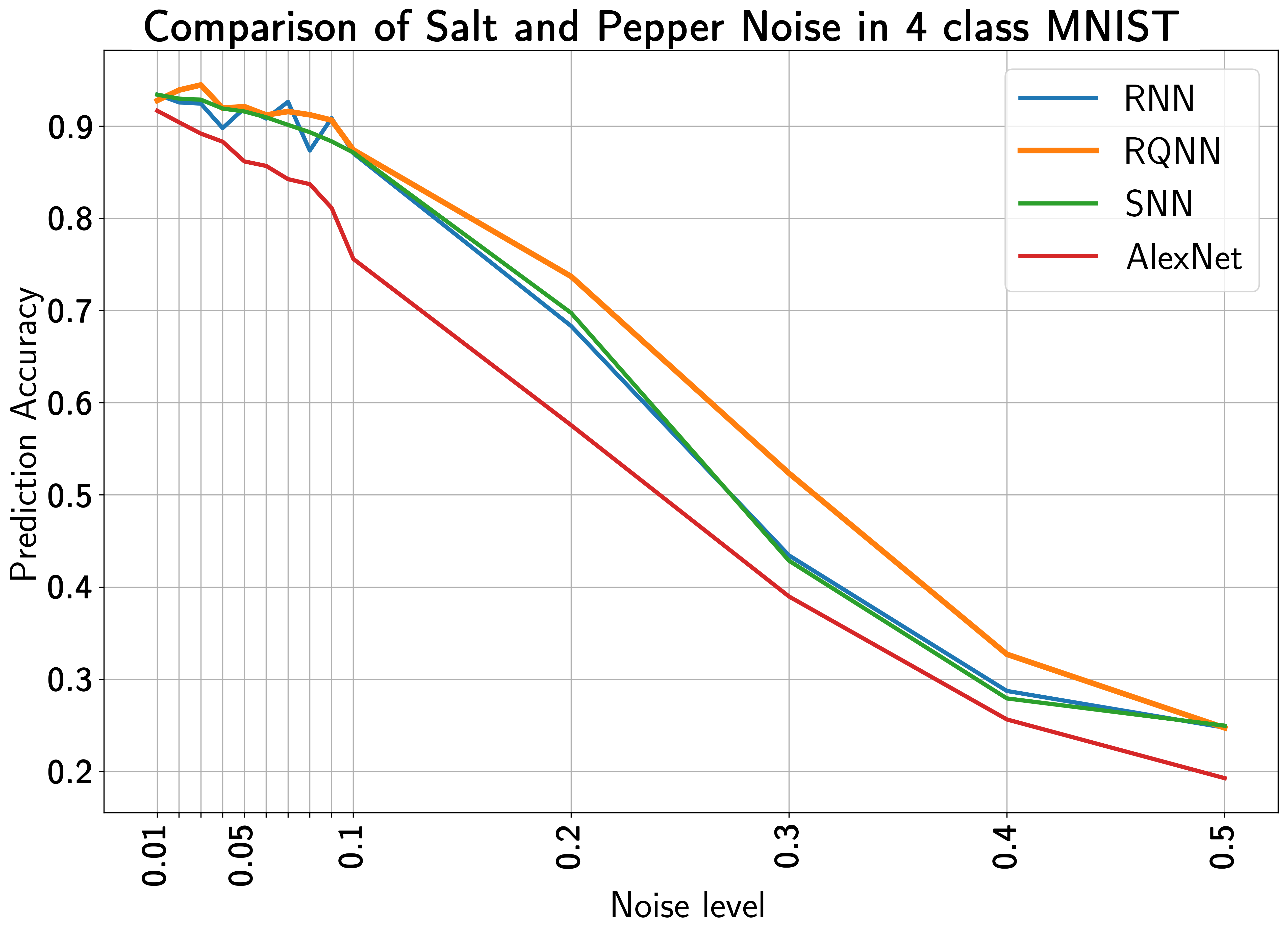}}
 \subcaptionbox{Rayleigh noise}{\includegraphics[width=2in]{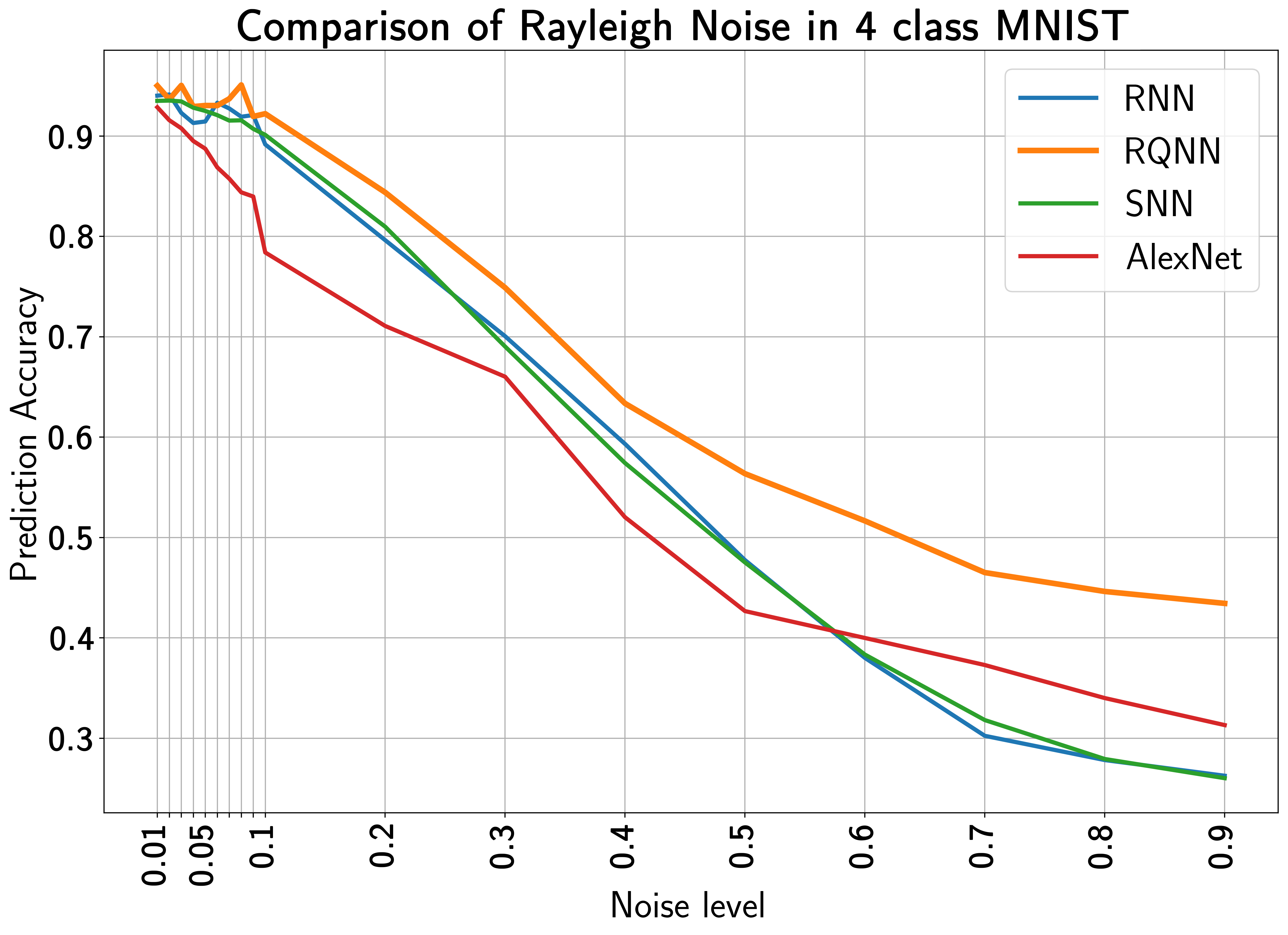}}
 \subcaptionbox{Uniform noise}{\includegraphics[width=2in]{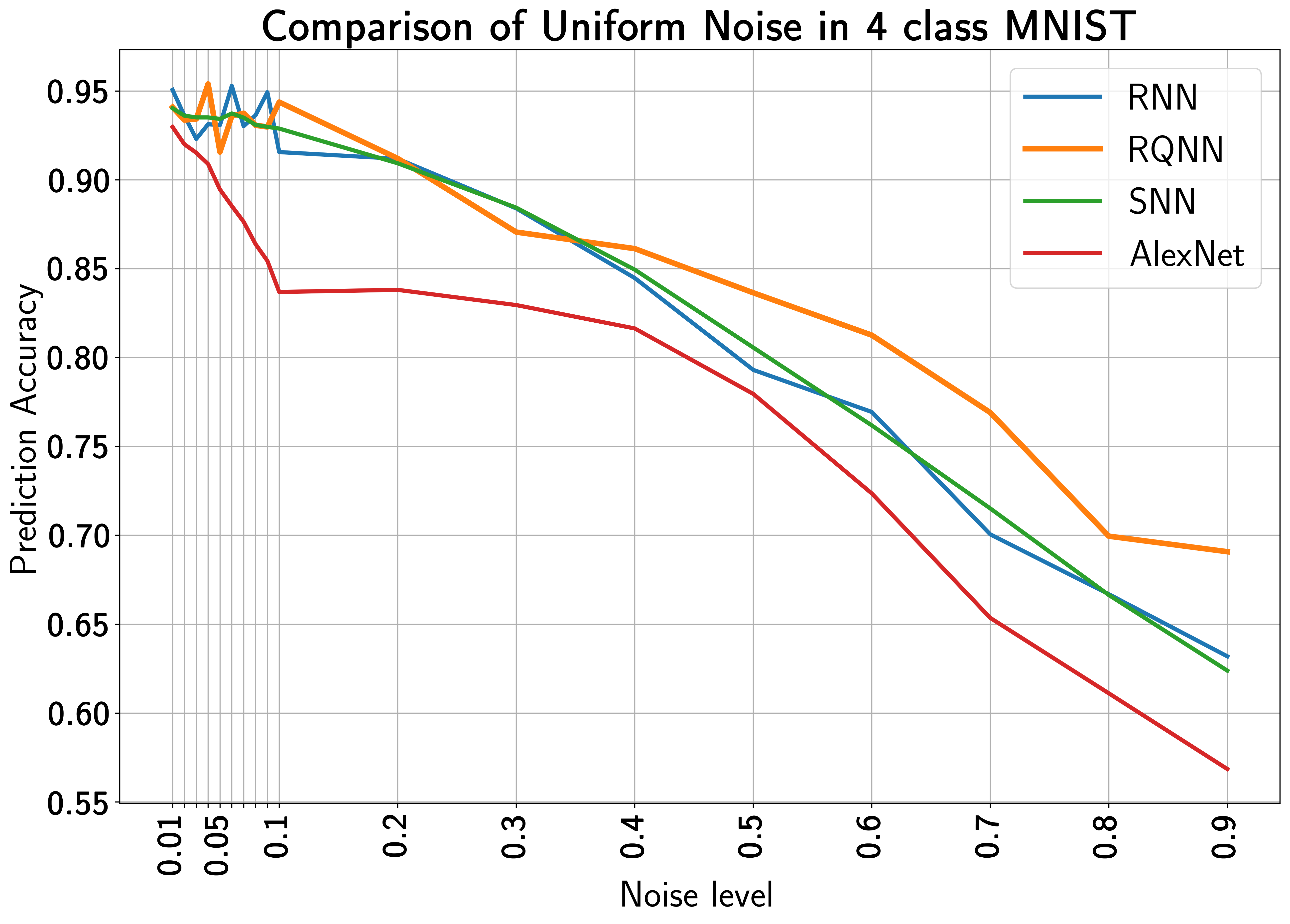}}
  \subcaptionbox{Perlin noise}{\includegraphics[width=2in]{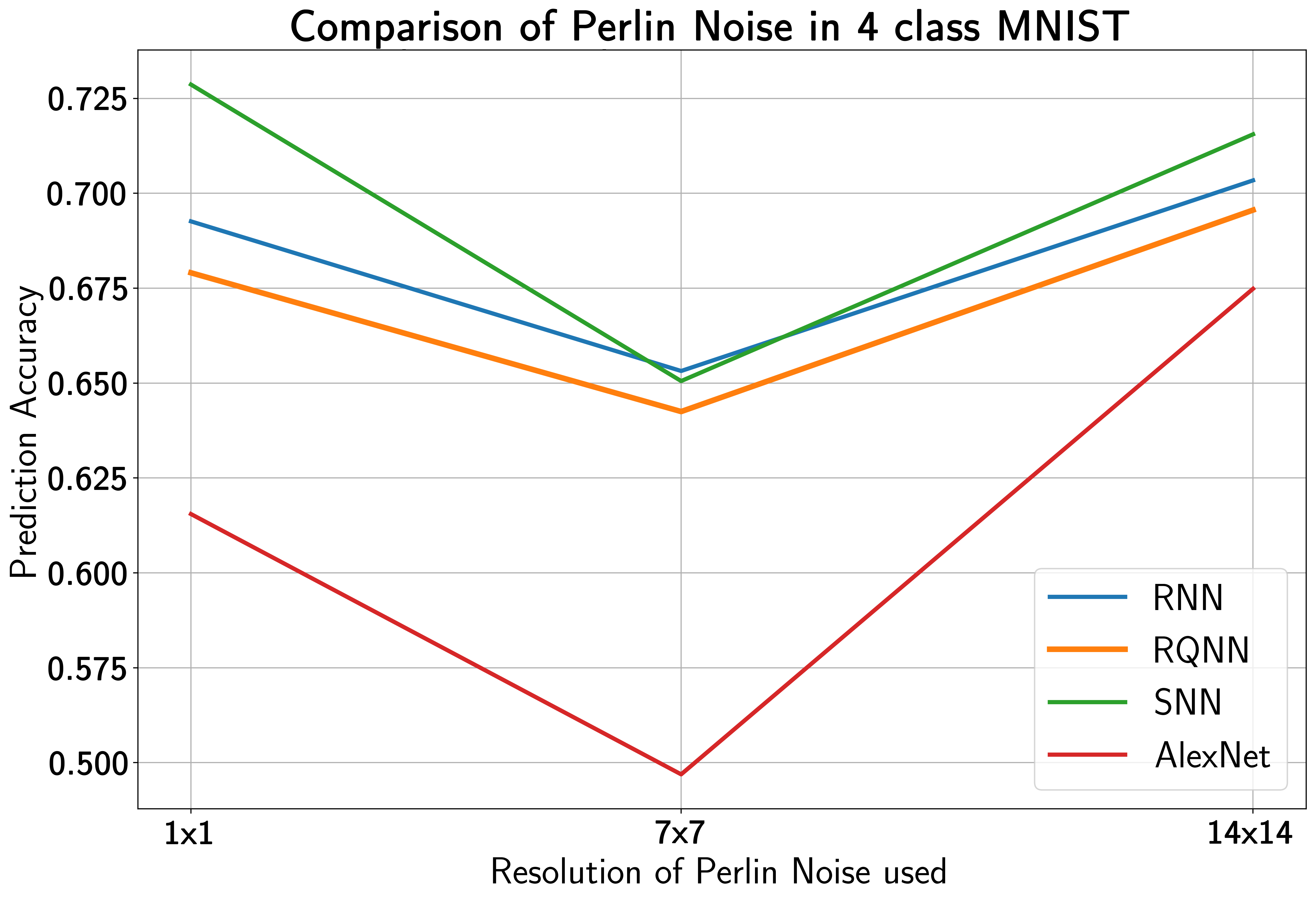}}
 \caption{Comparison of RQNN and classical RNN\cite{gelenbe2016} test accuracy for first four classes of MNIST dataset\cite{lecun1998} with (a) Gaussian noise (b) salt and pepper noise (c) Rayleigh noise (d) uniform noise and (e) Perlin noise.}
 	\label{fig:MNIST}
 \end{figure}

\begin{figure}[htbp]
	\centering
 \subcaptionbox{Gaussian noise}{\includegraphics[width=2in]{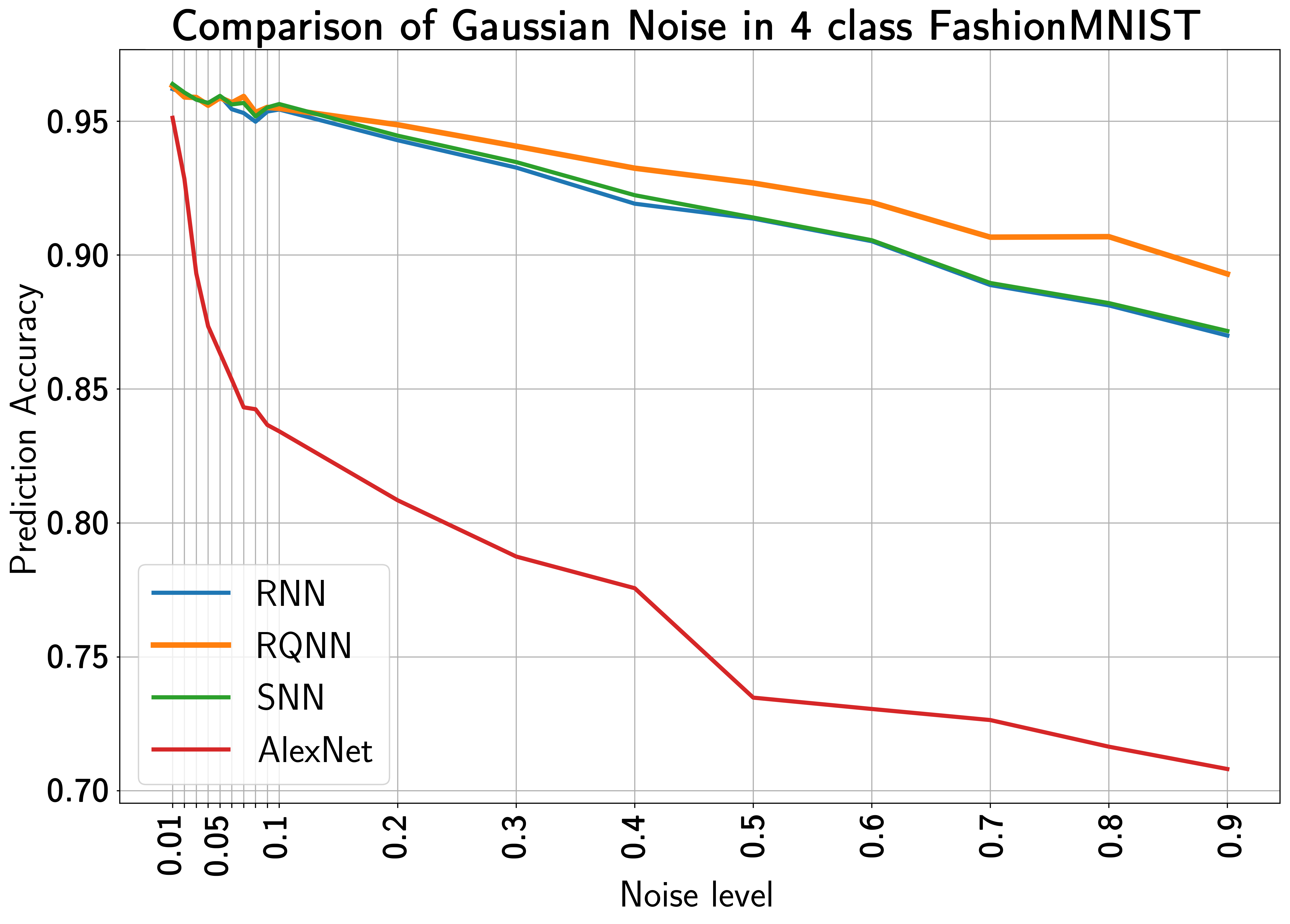}}
 \subcaptionbox{Salt and Pepper noise}{\includegraphics[width=2in]{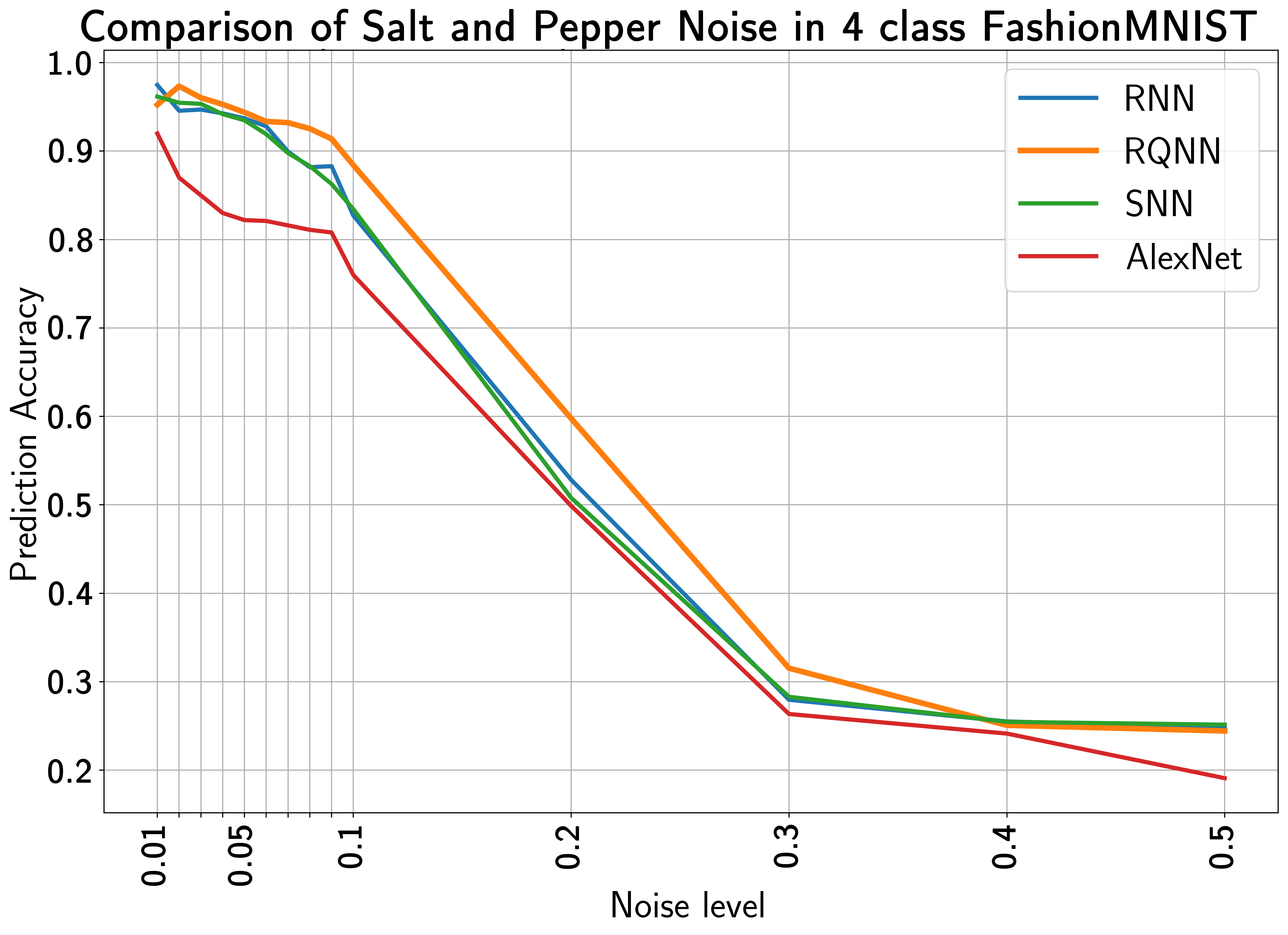}}
 \subcaptionbox{Rayleigh noise}{\includegraphics[width=2in]{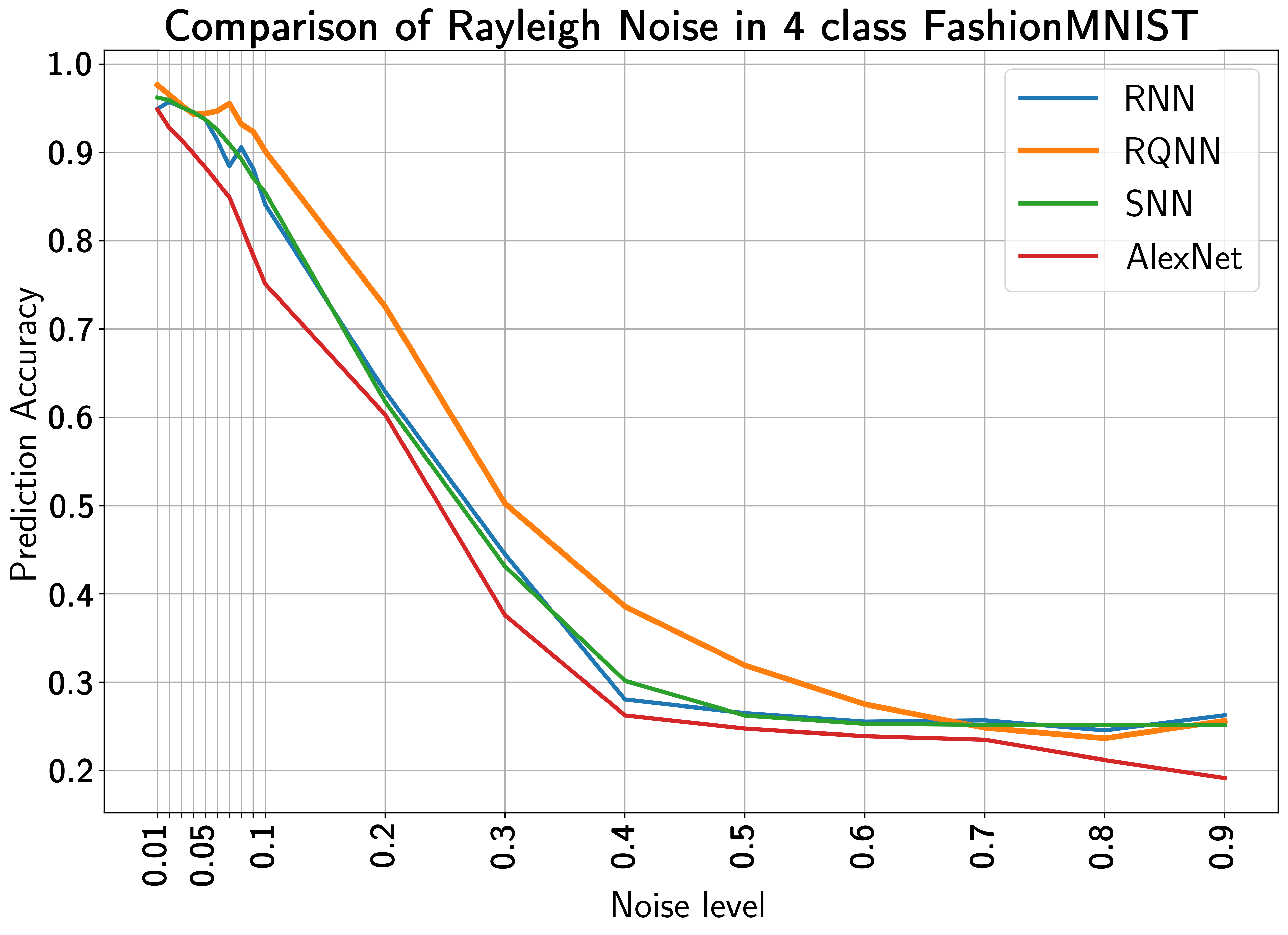}}
 \subcaptionbox{Uniform noise}{\includegraphics[width=2in]{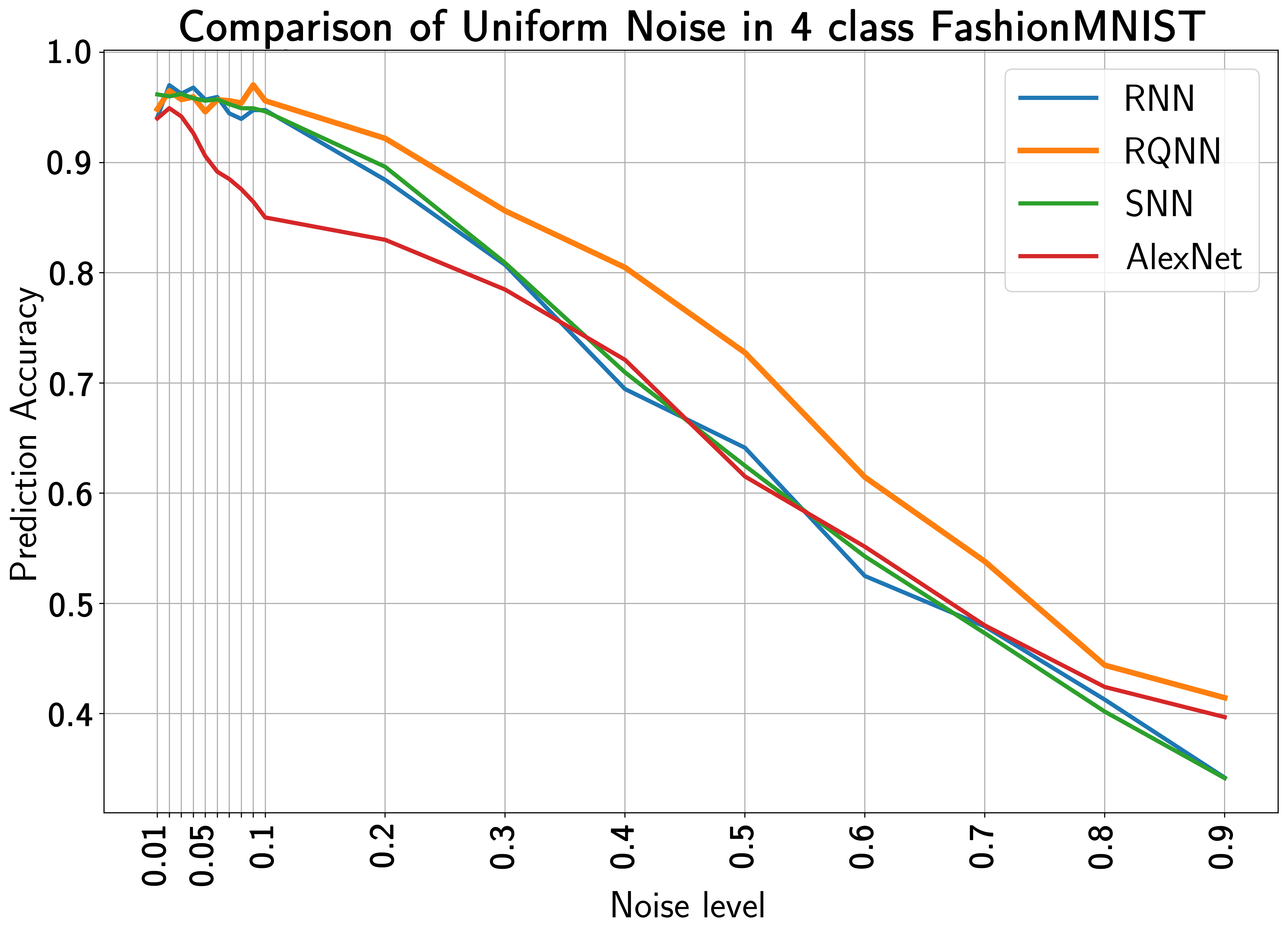}}
 \subcaptionbox{Perlin noise}{\includegraphics[width=2in]{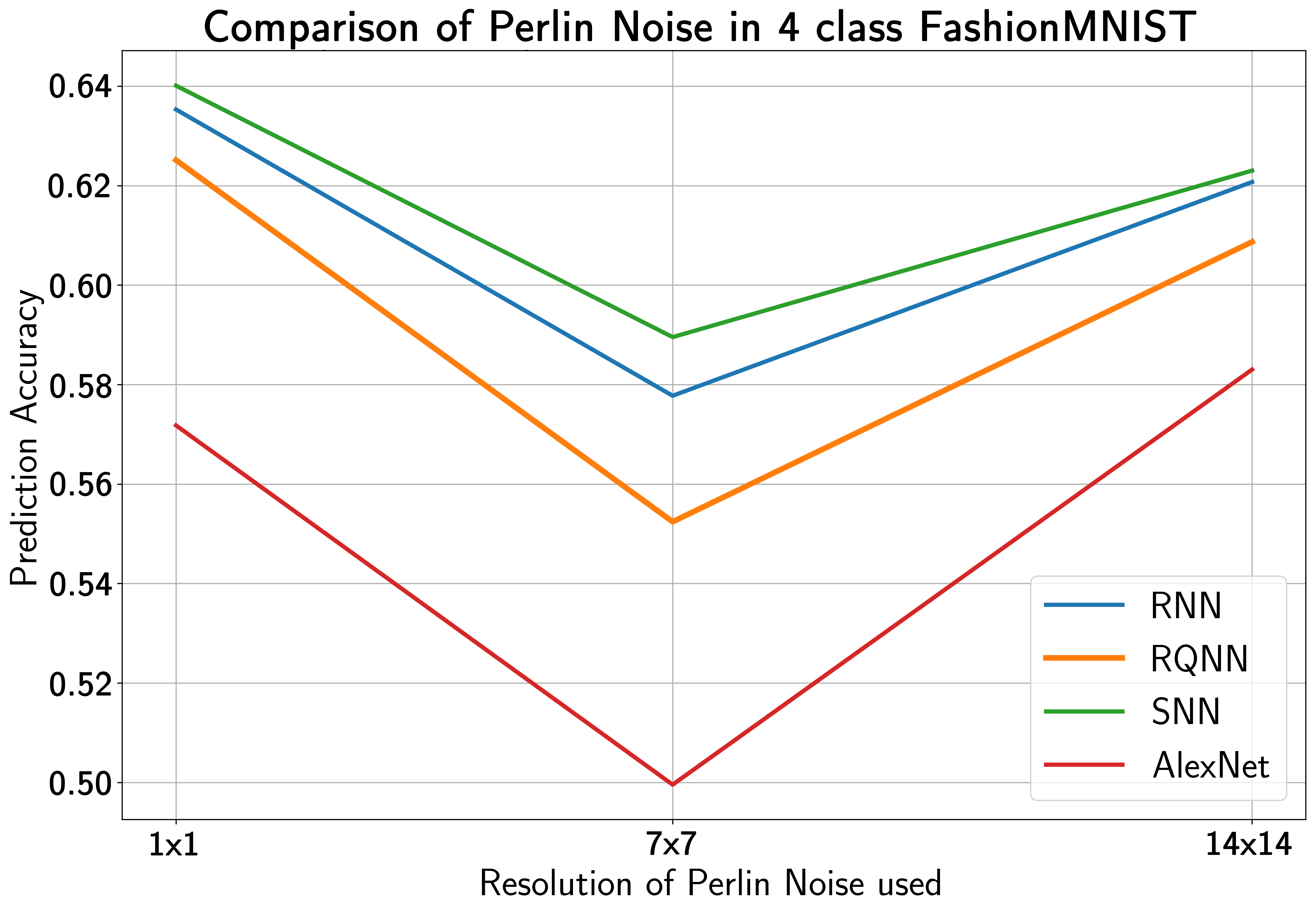}}
 \caption{Comparison of RQNN and classical RNN\cite{gelenbe2016} test accuracy for four classes of FashionMNIST dataset\cite{xiao2017} with (a) Gaussian noise (b) salt and pepper noise (c) Rayleigh noise (d) uniform noise and (e) Perlin noise.}
 	\label{fig:FashionMNIST}
 \end{figure}

Using the two-sided paired Wilcoxon signed-rank test, a non-parametric statistical hypothesis test, the proposed RQNN is compared to the state-of-the-art neural networks (classical RNN model\cite{gelenbe2016}, classical SNN\cite{khera2018}, and AlexNet\cite{alexnet2012}). The suggested RQNN model produces statistically significant outcomes in noisy settings, as evidenced by the Wilcoxon signed-rank test. In the case of Perlin noise, classical RNN and classical SNN marginally outperform our proposed RQNN due to the inherent repeatable pseudo-random values of the noise within a specified range. However, the overall accuracy of RQNN is significantly greater than that of other state-of-the-art methods, as shown in Table~\ref{tab2}. Figures~\ref{fig:MNIST} and \ref{fig:FashionMNIST} illustrate the mean accuracy values of noisy image recognition using the proposed RQNN and its classical counterparts with varied noise intensity levels. Figures~\ref{fig:MNIST} and \ref{fig:FashionMNIST} indicate that the proposed hybrid classical-quantum neural network (RQNN) outperforms its classical counterpart (RNN\cite{gelenbe2016}) in most noise types, as seen in the mean accuracy plots. 

\begin{table}
\tiny
	\begin{center}
		\caption{Analysis of the proposed RQNN compared to the classical RNN model\cite{gelenbe2016}, the classical SNN\cite{khera2018} and AlexNet\cite{alexnet2012} on test MNIST\cite{lecun1998}, FashionMNIST (FMNIST)\cite{xiao2017} and KMNIST\cite{clanuwat2018} datasets affected by salt and pepper (SnP), Gaussian, Rayleigh, uniform and Perlin noise [when comparing the findings with each component of the proposed RQNN, the bold values highlight the metrics having a $p$-value of $<0.05$ determined using the two-sided paired Wilcoxon signed-rank test.]}
	\begin{tabular}{p{40pt}p{30pt}p{25pt}p{15pt}p{15pt}p{15pt}p{15pt}}
			\hline
			\multirow{1}{*}{\centering{\textbf{Model}}} &
			\multirow{1}{*}{\centering{\textbf{Dataset}}} & \multirow{1}{*}{\centering{\textbf{Noise}}} & $\textbf{ACC}$ & $\textbf{DSC}$ & $\textbf{PPV}$ & $\textbf{SS}$ \\
			\hline
			\multirow{15}{*}{\textbf{RQNN}} & \multirow{5}{*}{MNIST} & SnP & $\textbf{0.965}$ & $\textbf{0.939}$ & $\textbf{0.985}$ & $\textbf{0.888}$ \\ 
			& &  Gaussian & $\textbf{0.969}$ & $\textbf{0.953}$	& $\textbf{0.990}$ & $\textbf{0.924}$\\	
			& &  Rayleigh & $\textbf{0.961}$ & $\textbf{0.904}$ & $\textbf{0.996}$ & $\textbf{0.852}$ \\
			& &  Uniform & $\textbf{0.972}$ & $\textbf{0.950}$ & $\textbf{0.988}$ & $\textbf{0.901}$ \\
			& &  Perlin & $0.927$	& $\textbf{0.841}$ & $0.882$ & $\textbf{0.813}$ \\
			\cline{2-7}
			 & \multirow{5}{*}{FMNIST} & SnP & $\textbf{0.915}$ & $\textbf{0.851}$ & $\textbf{0.927}$ & $\textbf{0.775}$ \\
			& &  Gaussian & $\textbf{0.968}$ & $\textbf{0.967}$	& $\textbf{0.962}$ & $\textbf{0.960}$\\	
			& &  Rayleigh & $\textbf{0.971}$ & $\textbf{0.942}$ & $\textbf{0.934}$ & $\textbf{0.922}$ \\
			& &  Uniform & $\textbf{0.975}$ & $\textbf{0.938}$ & $\textbf{0.974}$ & $\textbf{0.908}$ \\
			& &  Perlin & $0.833$	& $0.690$ & $0.713$ & $0.606$ \\
			\cline{2-7}
			& \multirow{5}{*}{KMNIST} & SnP & $\textbf{0.973}$ & $\textbf{0.710}$ & $\textbf{0.885}$ &	$\textbf{0.608}$ \\
			& &  Gaussian & $\textbf{0.979}$ & $\textbf{0.891}$	& $0.929$ & $\textbf{0.855}$\\
			& &  Rayleigh & $\textbf{0.965}$	& $\textbf{0.291}$ & $\textbf{0.966}$ & $0.165$ \\
			& &  Uniform & $\textbf{0.976}$	& $\textbf{0.783}$ & $\textbf{0.928}$ & $0.677$ \\
			& &  Perlin & $0.895$	& $0.309$ & $0.222$ & $0.531$ \\
			\hline
			\multirow{15}{*}{\textbf{RNN}} & \multirow{5}{*}{MNIST} & SnP & $\textbf{0.959}$ & $0.925$ & $0.971$ & $0.868$ \\ 
			& &  Gaussian & $\textbf{0.971}$ & $\textbf{0.956}$	& $0.974$ & $0.908$\\	
			& &  Rayleigh & $0.941$ & $\textbf{0.903}$ & $0.977$ & $0.827$ \\
			& &  Uniform & $0.953$ & $0.914$ & $\textbf{0.986}$ & $\textbf{0.891}$ \\
			& &  Perlin & $\textbf{0.933}$	& $\textbf{0.840}$ & $\textbf{0.893}$ & $0.802$ \\
			\cline{2-7}
			& \multirow{5}{*}{FMNIST} & SnP & $0.910$ & $0.812$ &	$0.902$ &	$0.723$ \\
			& &  Gaussian & $0.959$	& $0.947$ & $0.939$ & $0.941$\\	
			& &  Rayleigh & $0.954$	& $0.921$ & $0.913$ & $0.907$ \\
			& &  Uniform & $0.963$ & $0.931$ & $\textbf{0.973}$ & $0.888$ \\
			& &  Perlin & $0.852$ & $0.703$ & $0.732$ & $0.659$ \\
			\cline{2-7}
			& \multirow{5}{*}{KMNIST} & SnP & $\textbf{0.971}$ & $0.704$ & $\textbf{0.882}$ &	$0.598$ \\
			& &  Gaussian & $\textbf{0.979}$ & $0.882$	& $\textbf{0.934}$ & $\textbf{0.853}$\\
			& &  Rayleigh & $\text{0.943}$	& $0.272$ & $\textbf{0.968}$ & $0.161$ \\
			& &  Uniform & $0.968$	& $\textbf{0.781}$ & $0.923$ & $\textbf{0.684}$ \\
			& &  Perlin & $\textbf{0.931}$ & $0.365$ & $0.253$ & $\textbf{0.572}$ \\
		    \hline
			\multirow{15}{*}{\textbf{SNN}} & \multirow{5}{*}{MNIST} & SnP & $0.955$ & $0.924$ & $0.960$ & $0.863$ \\ 
			& &  Gaussian & $0.959$ & $\textbf{0.950}$	& $0.968$ & $\textbf{0.927}$\\	
			& &  Rayleigh & $0.934$ & $0.901$ & $0.979$ & $0.827$ \\
			& &  Uniform & $0.957$ & $0.906$ & $0.977$ & $0.886$ \\
			& &  Perlin & $\textbf{0.929}$	& $\textbf{0.838}$ & $0.880$ & $\textbf{0.811}$ \\
			\cline{2-7}
			& \multirow{5}{*}{FMNIST} & SnP & $0.911$ & $0.819$ & $0.918$ &	$0.739$ \\
			& &  Gaussian & $0.961$	& $0.949$	& $0.950$ & $0.951$\\	
			& &  Rayleigh & $0.960$	& $0.918$ & $0.924$ & $0.909$ \\
			& &  Uniform & $\textbf{0.971}$ & $\textbf{0.940}$ & $\textbf{0.972}$ & $0.879$ \\
			& &  Perlin & $\textbf{0.858}$	& $\textbf{0.721}$ & $\textbf{0.770}$ & $\textbf{0.679}$ \\
			\cline{2-7}
			& \multirow{5}{*}{KMNIST} & SnP & $0.957$ & $0.701$ & $0.880$ &	$0.575$ \\
			& &  Gaussian & $0.968$ & $0.857$ & $0.919$ & $0.846$\\
			& &  Rayleigh & $\text{0.938}$	& $\textbf{0.292}$ & $0.947$ & $\textbf{0.184}$ \\
			& &  Uniform & $0.957$	& $\textbf{0.780}$ & $0.906$ & $\textbf{0.681}$ \\
			& &  Perlin & $0.918$ & $\textbf{0.379}$ & $\textbf{0.271}$ & $0.564$ \\
			\hline
			\multirow{15}{*}{\textbf{AlexNet}} & \multirow{5}{*}{MNIST} & SnP & $0.888$ & $0.832$ & $0.821$ & $0.745$ \\ 
			& &  Gaussian & $0.758$ & $0.810$	& $0.761$ & $0.819$\\	
			& &  Rayleigh & $0.789$ & $0.832$ & $0.759$ & $0.682$ \\
			& &  Uniform & $0.787$ & $0.789$ & $0.820$ & $0.713$ \\
			& &  Perlin & $0.812$	& $0.641$ & $0.688$ & $0.712$ \\
			\cline{2-7}
			& \multirow{5}{*}{FMNIST} & SnP & $0.819$ & $0.748$ & $0.879$ & $0.726$ \\
			& &  Gaussian & $0.811$	& $0.923$	& $0.915$ & $0.912$\\	
			& &  Rayleigh & $0.802$	& $0.889$ & $0.846$ & $0.817$ \\
			& &  Uniform & $0.857$ & $0.889$ & $0.828$ & $0.736$ \\
			& &  Perlin & $0.779$	& $0.608$ & $0.616$ & $0.591$ \\
			\cline{2-7}
			& \multirow{5}{*}{KMNIST} & SnP & $0.752$ & $0.574$ & $0.709$ &	$0.498$ \\
			& &  Gaussian & $0.758$ & $0.626$ & $0.733$ & $0.563$\\
			& &  Rayleigh & $0.712$	& $0.201$ & $0.793$ & $0.127$ \\
			& &  Uniform & $0.722$	& $0.678$ & $0.734$ & $0.521$ \\
			& &  Perlin & $0.728$ & $0.315$ & $0.226$ & $0.534$ \\
			\hline
		    \label{tab2}
		\end{tabular}
	\end{center}
\end{table}
\begin{figure}[htbp]
	\centering
 \subcaptionbox{Gaussian noise}{\includegraphics[width=2in]{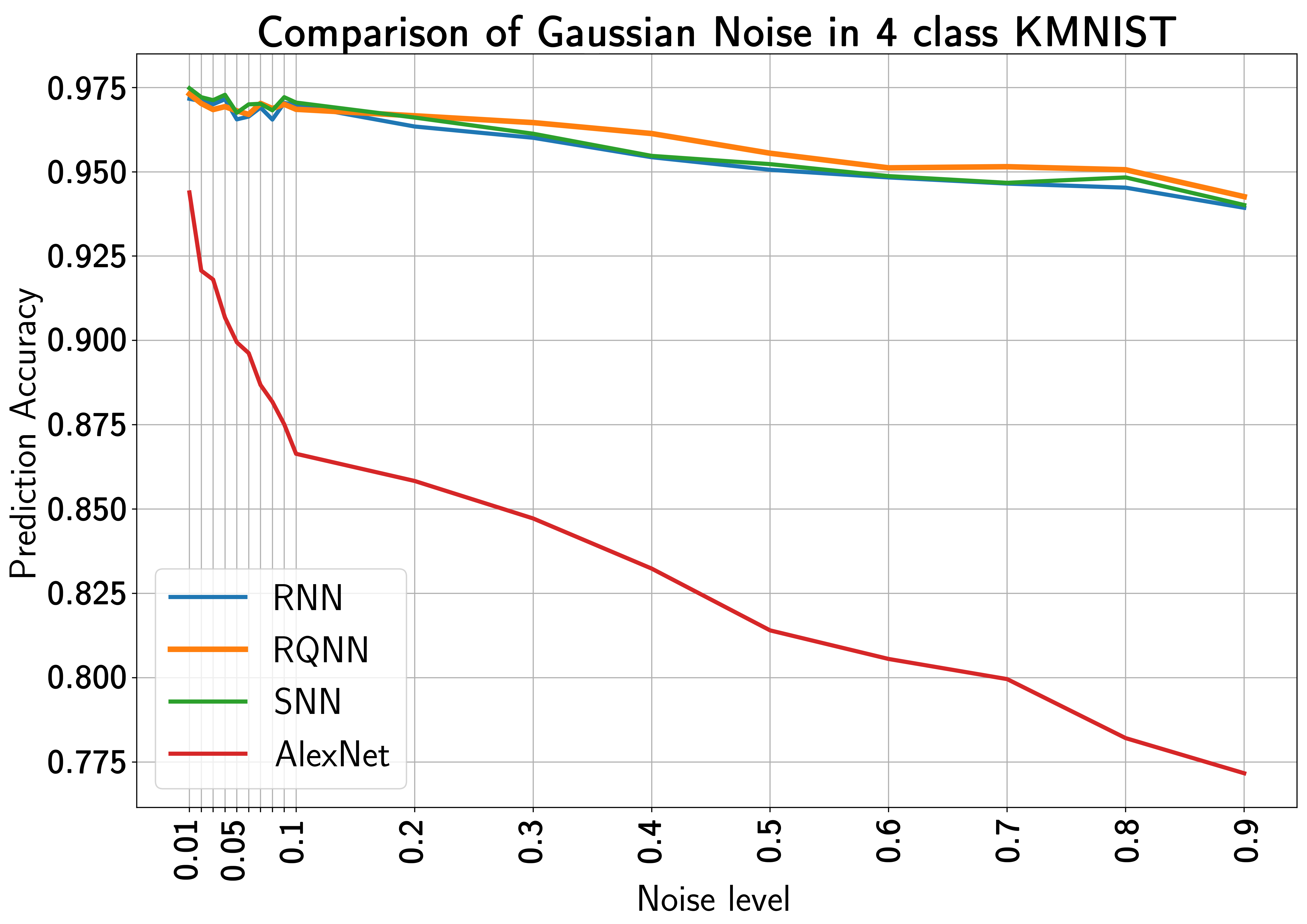}}
 \subcaptionbox{Salt and Pepper noise}{\includegraphics[width=2in]{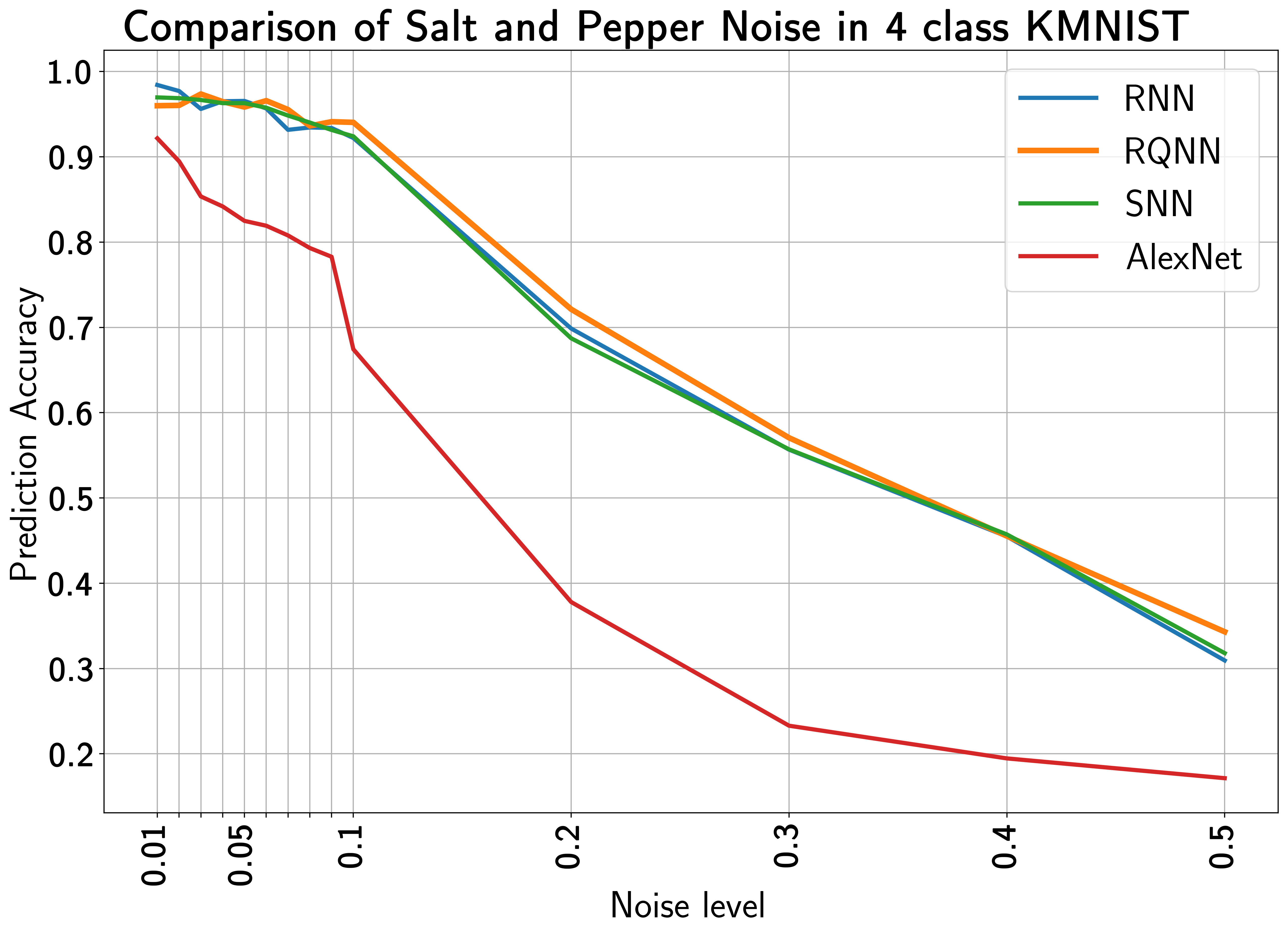}}
 \subcaptionbox{Rayleigh noise}{\includegraphics[width=2in]{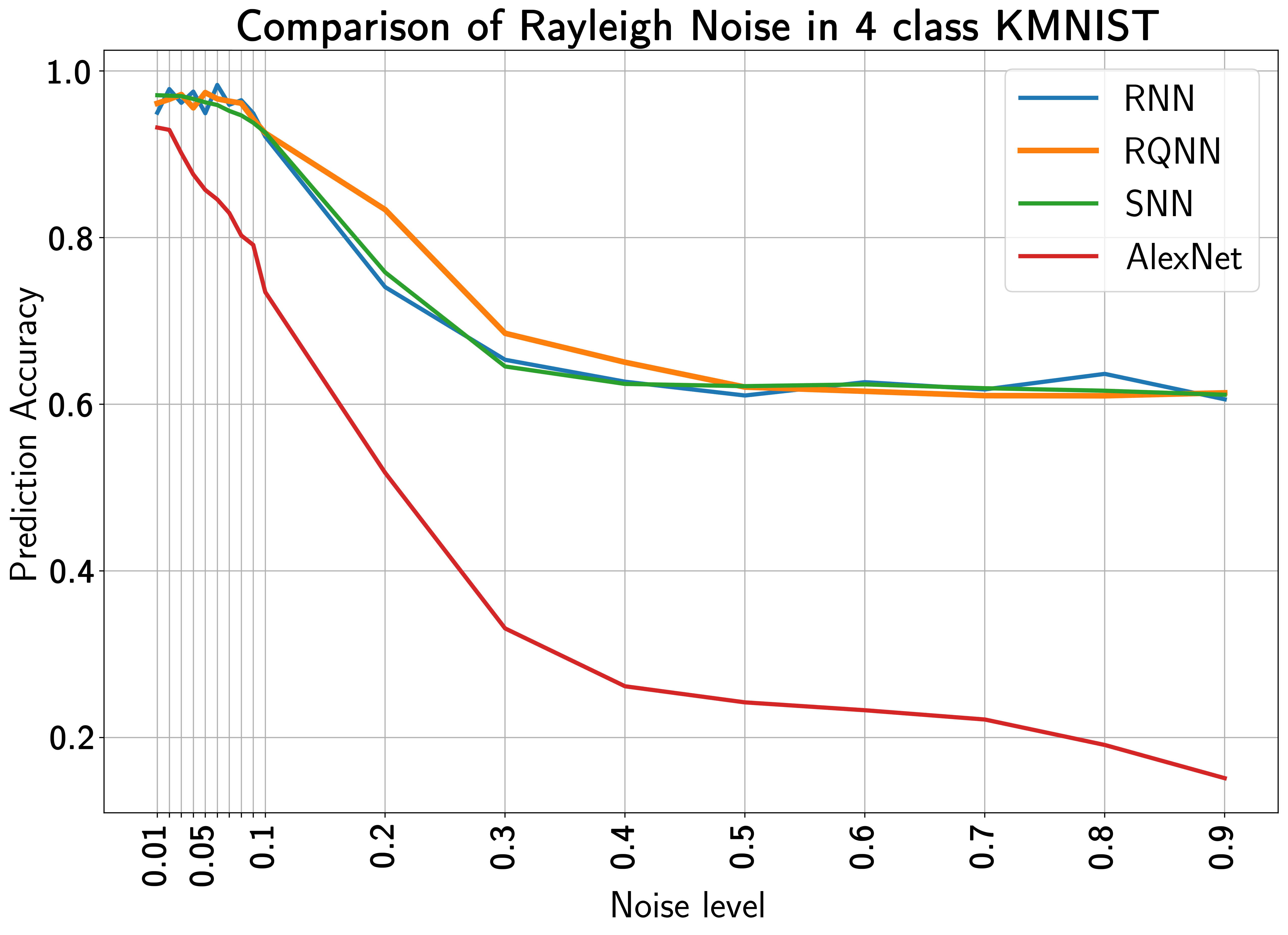}}
 \subcaptionbox{Uniform noise}{\includegraphics[width=2in]{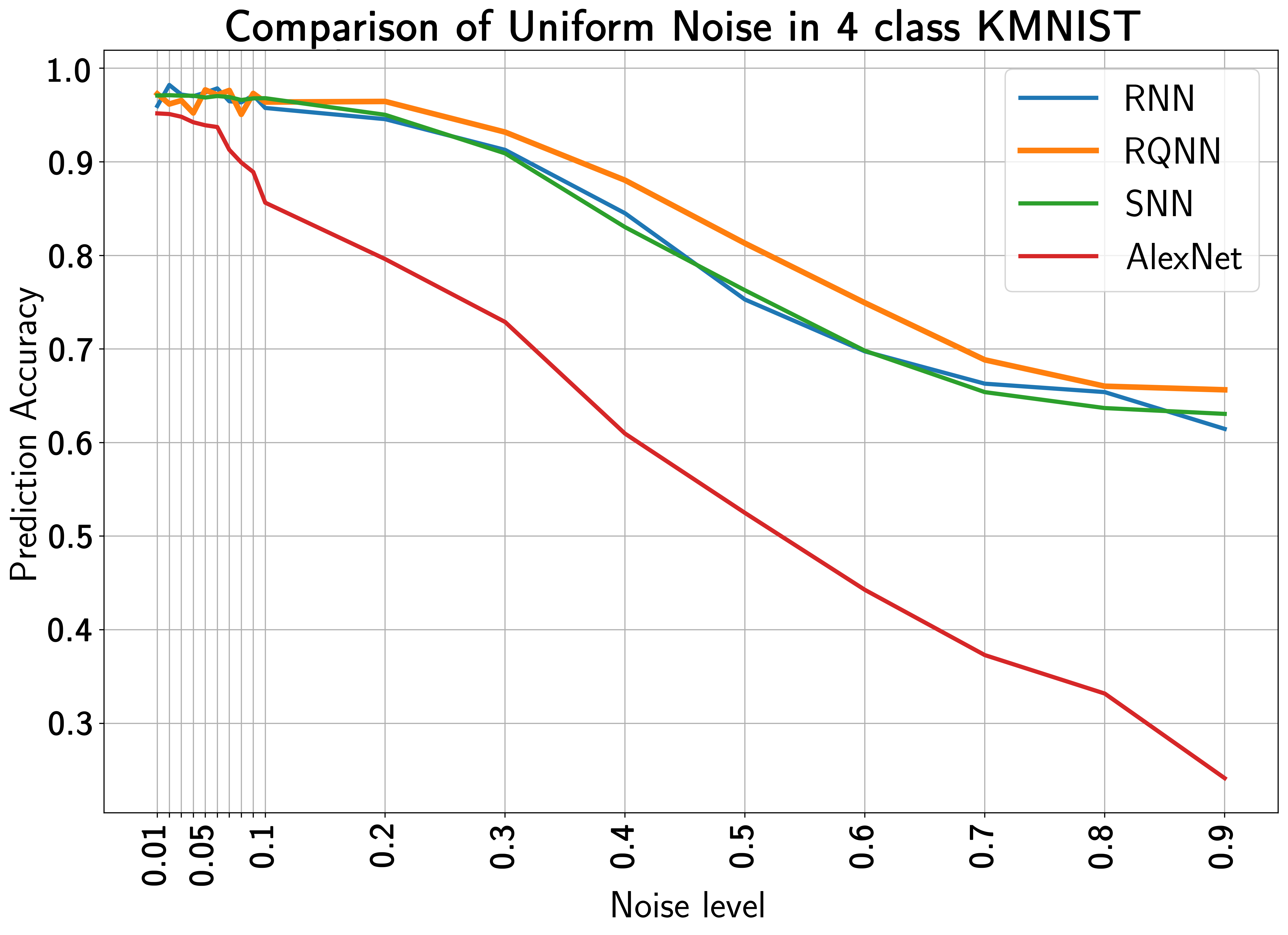}}
  \subcaptionbox{Perlin noise}{\includegraphics[width=2in]{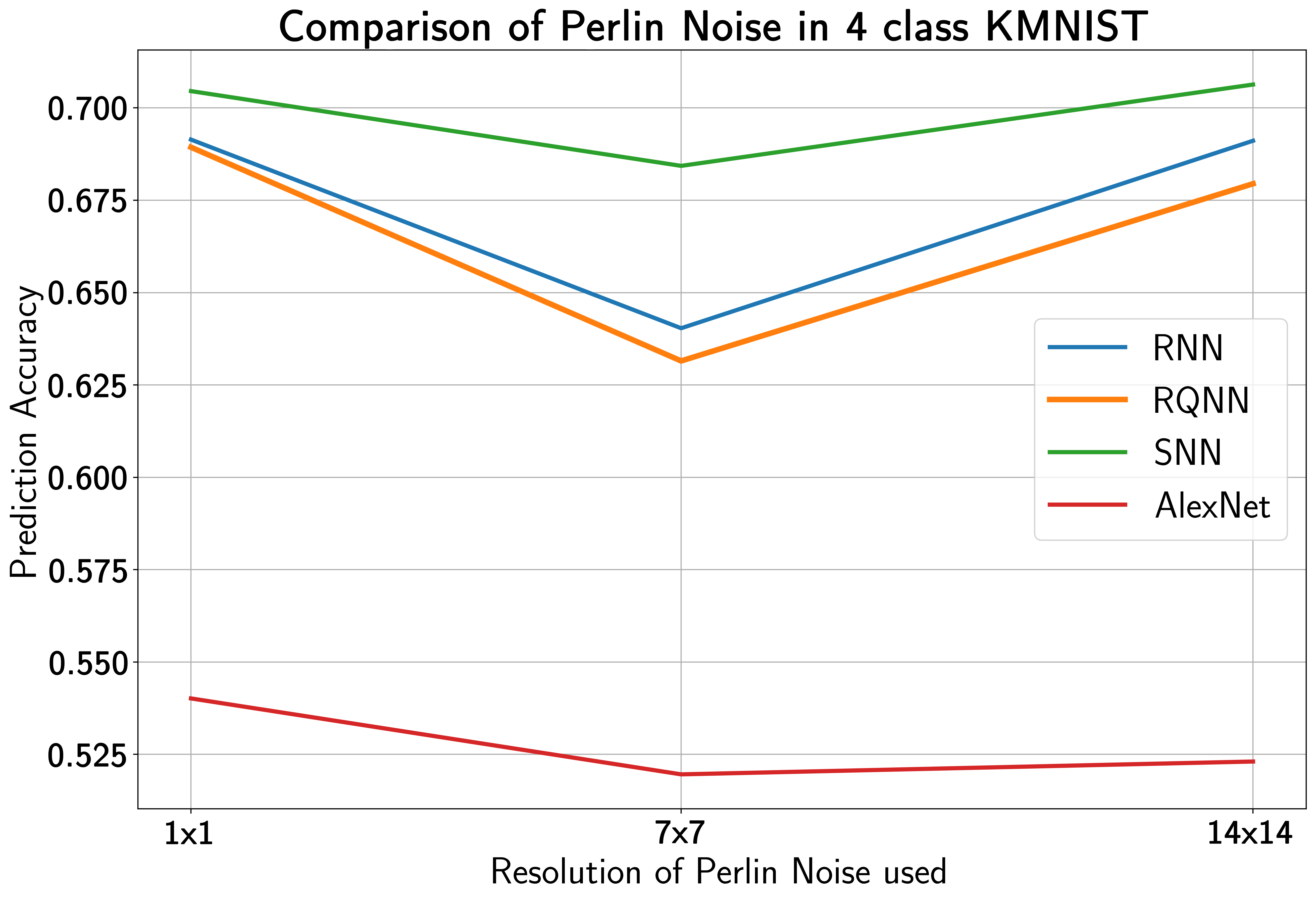}}
 \caption{Comparison of RQNN and classical RNN\cite{gelenbe2016} test accuracy for first four classes of Kuzushi-49 dataset\cite{clanuwat2018} with (a) Gaussian noise (b) salt and pepper noise (c) Rayleigh noise (d) uniform noise and (e) Perlin noise.}
 	\label{fig:Kuzushi-49}
 \end{figure}
 
\section*{Discussion}
\label{discuss}

We developed a novel class of supervised RQNNs based on a hybrid classical-quantum algorithm. The RQNN exhibits superior accuracy over conventional neural networks in noisy environments. The suggested RQNN model outperforms its classical counterpart (RNN)\cite{gelenbe2016}, classical SNNs\cite{khera2018}, and the AlexNet model\cite{alexnet2012}. It is simply because the classical RNNs\cite{gelenbe2016}, classical SNN\cite{khera2018}, and CNNs (AlexNet)\cite{alexnet2012} struggle to modify their network weights in noisy settings. Due to the intrinsically probabilistic character of qubits, the dressed quantum layer in the proposed RQNN efficiently approximates the noise with original pixel values, resulting in improved classification accuracy. Furthermore, the dressed quantum layer in the proposed RQNN model can introduce entanglement between the inputs and outputs of the random spiking neurons. The input state can be influenced in extremely non-trivial ways due to measurement back-action from typical output measurements.

More specifically, the RQNN is trained on smaller datasets to optimize each parameter in the VQC at different epochs. In contrast, all the parameters in classical deep learning models can be updated at every single epoch in parallel. While the RQNN yields a superior performance over classical RNN, SNN, and AlexNet on smaller datasets with similar training sets, RQNN is limited by the inherent challenges in the scalability of VQCs. However, we demonstrated here that our proposed RQNN framework yields superior accuracy despite the current technical limitations on the number of available qubits. Furthermore, we devise an efficient optimization method for random spiking quantum neurons by leveraging the intrinsic uncertainty in qubits by utilizing hybrid classical-quantum algorithms. Experiments employing the proposed RQNN model on the noisy MNIST, FashionMNIST and KMNIST datasets indicate that it outperforms its classical equivalents and deep learning models (AlexNet). The RQNN achieves recognition accuracy of over $94.9$ on the standard problem of visual character recognition on very large datasets. The quality of the observed results appears to improve as the network grows in size with the number of qubits. Hence, the proposed RQNN model appears to be a promising candidate for various computer vision applications and has the potential to revolutionize quantum machine learning research. However, it remains to be investigated how the proposed RQNN architecture performs in other computer vision applications. The proposed RQNN model's robust experimental findings in handling noise with RQNN and its superior performance with respect to state-of-the-art classical RNN models stimulate future applications in various applications such as pattern recognition, optimization, learning, and associative memory. Furthermore, the proposed RQNN model incorporates hybrid classical-quantum algorithms, which improves network accuracy in a noisy environment. This novel hybrid classical-quantum spiking random neural network is therefore suitable for implementation on near-term quantum devices straight away.

Finally, in this study, the hyper-parameters of the proposed RQNN model are not fine-tuned to attain similar accuracy to CNN models for image classification without noise. Despite this fact, the RQNN outperformed the RNN\cite{gelenbe2016}, classical SNN\cite{khera2018}, and AlexNet\cite{alexnet2012} in terms of accuracy ($ACC$), dice similarity ($DS$), positive predictive value ($PPV$), and sensitivity ($SS$) for unseen noisy test images. The suggested model is statistically significant and could be a viable candidate for deep learning algorithms for other computer vision challenges in the future. It is interesting to note that the proposed RQNN model requires fewer parameters to implement than its classical equivalents, thanks to the parametrized VQC in the dressed quantum layer, which decreases the number of training parameters. Nonetheless, we aim to build an RQNN model to optimize hybrid classical-quantum circuits with millions of parameters relatively efficiently (a serious problem for classical RNNs). It is also worth noting that in this work, the hybrid classical-quantum circuit is the leading contender for unitary random neural networks.

\section*{Methods}
\label{Methods:RQNN}
\subsection{Hybrid Classical-Quantum Random Neural Network Architecture.}
\label{Architecture:RQNN}

The proposed hybrid classical-quantum RQNN model consists of a classical RNN\cite{gelenbe2016, gelenbe2020}, a dressed quantum layer for classification (whose qubits are further measured to collapse to classical bits), and a final classical post-processing layer for determining the results. In addition, as illustrated in Figure~\ref{fig:RQNN}, a classical preprocessing layer is used as a connecting layer between the temporal pooling layer and the dressed quantum layer of the model design. The classical multi-layer random neural network processes the input data with input dimensions exceeding the number of available qubits. It reduces the high-dimensional images to low-dimensional feature representations. However, one of the uphill tasks in quantum machine learning in the NISQ era is to efficiently encode high-dimensional classical bits into qubits in the dressed quantum layer. In our hybrid classical-quantum architecture, the classical RNN's\cite{gelenbe2016, gelenbe2020} intermediate outputs are encoded as quantum states and sent to the adjacent VQC. A brief introduction to the authors' previously invented classical RNN models\cite{gelenbe2016, gelenbe2020} is provided in the following section.

\subsection{Classical Random Neural Networks (RNN).} 
\label{Architecture:RNN}
The conventional RNNs\cite{gelenbe2016, gelenbe2020} are composed of neurons that receive excitatory (positive) and inhibitory (negative) spike impulses from external sources, sensory sources, or neurons. At time $t=0$, each neuron's internal state is designated as $\alpha_k (t)\geqslant 0$, \emph{i.e.}, a non-negative integer. These spike signals can arrive from other neurons, since each neuron which is excited, \emph{i.e.}, whose internal state $\alpha_k(t) > 0$ will fire after an exponentially distributed random time of parameter $\mu_k$, or are generated separately by external Poisson processes with rates of $\lambda_+ (k)$ for excitatory spike signals and $\lambda_- (k)$ for inhibitory spike signals to a neuron $k$. If $\alpha_k (t) = 0$, then a negative spike to neuron $k$ at time $t$ has no effect on its internal state since $\alpha_k\geqslant 0$.
\begin{equation}
  \alpha_k (t^+) = \max \{\alpha_k (t) -1, 0\}
\end{equation}
The arrival of an excitatory spike, on the other hand, always increases the neuron's internal state by one as follows.
\begin{equation}
    \alpha_k (t^+) = \alpha_k (t) +1
\end{equation}
When neuron $k$ "fires", which occurs after an exponentially distributed delay of parameter $\mu_k\geqslant 0$ for any neuron $k$,  the signal either travels toward neuron $j$ with probability $p^+ (k,j)$ as a positive signal, or travels towards neuron $j$ with probability $p^- (k, j)$ as a negative signal, or exits the network with probability $d(k)$. The transition probability of the Markov chain reflecting the flow of signals between neurons is as follows: 
\begin{equation}
 p(k, j) = p^+(k, j) + p^- (k,j) 
\end{equation}
and
\begin{equation}
 \sum_j p(k,j) + d(k) = 1, \forall 1 \leq k \leq n 
\end{equation}
where $n$ is the RNN model's number of neurons. \\
If neuron $k$ is excited (i.e. $\alpha_k (t) > 0$), it will fire after an exponentially distributed time of parameter $\mu_k\geqslant 0$, and  its internal state decreases by one ($\alpha_k (t+\Delta t) = \alpha_k (t) -1$ with probability $\mu_k\Delta t$), and it has the probability $p^+ (k,j)$ of delivering a positive or excitatory spike to neuron $j$, resulting in $\alpha_j (t+\Delta t) = \alpha_j (t) +1$ with probability $\mu_k\Delta t p^+(k,j)$. Alternatively, it can transmit a negative or inhibitory spike to neuron $j$ with probability $p^-(k,j)$, resulting in $\alpha_j (t+\Delta t) = \alpha_j (t) -1$ if $\alpha_j (t) > 0$ and $\alpha_j (t) = 0$ if $\alpha_j (t) = 0$ with probability $\mu_k\Delta t p^-(k,j)$. RNNs can also contain ``triggers'' which allow firing to propagate simultaneously through several cells.

\subsection{Quantum Variational Circuit (VQC) of RQNN.} 

The major goal of this research is to present a universal RQNN model that employs qubits to provide a reasonably accurate and resilient classification of noisy images in hybrid classical-quantum settings. The quantum information processing on qubits using various quantum gate operations are detailed below. \\
The following is an illustration of a general quantum system with $D$ qubits\cite{konar2021}. 
\begin{equation}
  |\psi\rangle = \sum_{(\phi_1, \phi_2, \cdots \phi_D \in \{0, 1\} )} \lambda_{\phi_1, \phi_2, \ldots \phi_M} |\phi_1\rangle \otimes |\phi_2\rangle \cdots \otimes |\phi_D\rangle
\end{equation}
Where $\lambda_{\phi_1, \phi_2, \ldots ,\phi_D}$ are complex amplitudes of each \emph{qubit} in the quantum system $|\psi\rangle$ subject to the fulfilment of the following condition.
\begin{equation}
    |\lambda_{\phi_1}|^2 + |\lambda_{\phi_2}|^2 + \cdots +|\lambda_{\phi_D}|^2 =1
\end{equation}
A qubit system $|\psi\rangle$ of dimension $\log n$\cite{konar2016} can be generated by combining a collection of $n$ base states (represented as $|\phi_j$) containing $0-1$, as 
\begin{equation}
|\psi\rangle = \sum_j^n \lambda_{\phi_j} |\phi_j\rangle
\end{equation}
Where $\lambda_{\phi_j}$ designates the probability amplitude and $|\psi\rangle = |\phi_1\rangle \otimes |\phi_2\rangle\otimes \cdots |\phi_n\rangle$. In contrast to conventional bits, the coherent quantum state $|\phi\rangle$ exists in superposition with the eigenstates $|0\rangle$ and $|1\rangle$. The Born rule\cite{fenyman1965} is used to determine the probability of $|1\rangle$ in the eigenstate $|\phi\rangle$. We incorporated an amplitude encoding scheme to transform the classical bits into quantum bits or qubits\cite{havlicek2019}.
\begin{equation}
\begin{split}
(\mathcal{H}|\phi\rangle)^{\otimes D} =   \underbrace{\mathcal{H}|\phi\rangle \otimes \mathcal{H}|\phi\rangle \cdots \otimes \mathcal{H}|\phi\rangle}_\text{D}\\
= \underbrace{[\frac{1}{\sqrt{2}} (|0\rangle + |1\rangle)] \otimes \cdots \otimes [\frac{1}{\sqrt{2}} (|0\rangle + |1\rangle)]}_\text{D}\\
= [\frac{1}{\sqrt{2}} (|0\rangle + |1\rangle)]^{\otimes D} = \frac{1}{\sqrt{2^D}} (|0\rangle + |1\rangle)^{\otimes D} \\= \frac{1}{\sqrt{2^D}} (|0\rangle \otimes \cdots \otimes |0\rangle + \cdots + |1\rangle \otimes \cdots |1\rangle)
= \frac{1}{\sqrt{2}} \sum_{i=1}^D |i\rangle
\end{split}
\end{equation}
The VQC employed in this framework is divided into a trinity of sections: encoding, variational, and measurement. The encoding part consists of the Hadamard gate $H$ and single qubit rotation gates $\mathcal{R}_x(\omega)$ and $\mathcal{R}_z(\omega)$, representing rotations along the x-axis and z-axis, respectively. To produce an impartial starting state, the Hadamard gate $H$ is introduced. The input state of VQC in the dressed quantum layer is labelled in quantum bits (qubits) in this proposed RQNN.
\begin{equation}
    |\mathcal{X}(\omega)\rangle = (\cos \omega |0\rangle + \sin \omega |1\rangle)|x\rangle
\end{equation}
where, $|\mathcal{X}(\omega)\rangle$ is a quantum state in the RQNN model's dressed quantum layer corresponding to the classical information $|x\rangle$ from the classical layers of RNNs. The $|\mathcal{X}(\omega)\rangle$ quantum state or qubit characteristics are sent to a Hadamard gate, $\mathcal{H}$, for equal superposition of the qubit states\cite{konar2021}.
\begin{equation}
\mathcal{H} = \frac{1}{\sqrt{2}}\left[ {{\begin{array}{*{20}c}
		1 & ~1 \hfill \\
		1 & -1 \hfill \\
		\end{array} }} \right].
\end{equation}
The encoding phase, which comprises $\mathcal{R}_x(\omega)$ and $\mathcal{R}_z(\omega)$ rotations, is applied to this initial quantum state in the proposed VQC of the dressed quantum layer of RQNN. The CNOT gates are employed to create complementary quantum states. The $\mathcal{R}_x(\omega)$ and $\mathcal{R}_z(\omega)$ gates represent single-qubit rotations through angle $\omega$ (radians) around the $x$ and $z$-axes on the Bloch sphere projection, respectively. The angle of rotation is characterized by the inputs from the classical layer of the RNN, as shown in Figure~\ref{fig:RQNN}. Subsequently, the encoded state is processed using variational quantum circuits with parameter optimization.
\begin{equation}
    \mathcal{R}_x(\omega)= \exp{(-j\mathcal{X}\omega/2)}\left[ {{\begin{array}{*{20}c}
		{\cos\omega/2 } \hfill & {-j\sin\omega/2 } \hfill \\
		{\sin\omega/2 } \hfill & {~~\cos\omega/2 } \hfill \\
		\end{array} }} \right]
\end{equation}
and 
\begin{equation}
    \mathcal{R}_z(\omega)= \exp{(-j\mathcal{Z}\omega/2)}\left[ {{\begin{array}{*{20}c}
		{\exp{(-j\omega/2)}} \hfill & ~~0 \hfill \\
		~~0 \hfill & 	{\exp{(j\omega/2)}} \hfill \\
		\end{array} }} \right]
\end{equation}
The variational component of the proposed VQC, which includes CNOT gates for entangled quantum states from individual qubits as well as $\mathcal{R}_x(\omega)$ and $\mathcal{R}_z(\omega)$ parameters for the universal single qubit unitary gate with $16$ parameters to be learned ($p_0, p_1, \cdots p_{15}$). The weights in classical random neural networks can be compared to these circuit properties. In the VQC of the RQNN circuit, a single quasi-local unitary ($U_i$) gate is used in a transitionally invariant form for finite depth. The recovered data will next be processed classically using Softmax to calculate the probability of each potential class. The result is obtained using quantum observations or measurements on qubits. In the RQNN model, each qubit has a starting state (for example, an evenly balanced superposition of the two values, $0$ and $1$). The rotating gate increases the likelihood of receiving $1$ or $0$. An $n$-qubit vector probabilistically depicts the whole space of all the $2^n$ combinations of incoming spikes to neurons at a particular instance. In terms of space, classical encoding data in the amplitudes of a superposition is the most economical method in utilizing $n$ qubits to encode $2^n$-dimensional incoming spikes. Applying the rotation operator to all $n$ qubits is the same as running a parallel search for the best state in the space. Compared to classical deep spiking neural network models, O($log_2n$) parameters characterize a set of $n$ qubit-based dressed quantum layers of RQNN and exponentially lower the number of parameters\cite{gelenbe2016}. The training of the VQC relies on quantum back-propagation algorithms. Here, the number of VQC layers is derived by optimizing the circuits for optimal expressibility of the encoded qubits in Hilbert space. It aims to obviate “barren plateaus” problems of local minima\cite{havlicek2019}. The VQC's convergence study is available in the \emph{Appendix}.

\section*{Appendix}
\scriptsize
\subsection{Convergence Analysis of Random Quantum Neural Networks.} 
\label{PQIS-Net:Conv}
The loss function assists the dressed quantum layer of RQNN to converge, and the classification accuracy is gained after the network stabilizes. The RQNN adds the cross-entropy loss function for classification, which is evaluated as follows.
\begin{equation}
   \operatorname*{argmin}_\theta \mathcal{L}_{\theta} = \sum_{j}^{\mathcal{C}} [\gamma_j \log z (\omega_j) + (1-\gamma_j) \log \{1-z (\omega_j)\}]
\end{equation}
The true output is $\gamma_j$, but the intended result of the Fully Connected (FC) layers on the input $\omega_j$ for the network hyper-parameter set $theta$ is $z(\omega_j)$. For the rotation gates $\mathcal{R}_x$ and $\mathcal{R}_y$ of VQC in RQNN, the rotation angle (variational parameter ($\omega$) is $\rho$ and $\sigma$, respectively. The rotation gates $\mathcal{R}_x$ and $\mathcal{R}_y$ of VQC control the qubits $\phi (x)$ and $\phi (y)$, respectively.
\begin{equation}
|\phi (x)^{\vartheta +1}\rangle=\left(
\begin{array}{cc}
\cos \triangle \rho^\vartheta  & -\sin \triangle \rho^\vartheta  \\
\sin \triangle \rho^\vartheta  & \cos \triangle \rho^\vartheta \\
\end{array}
\right)|\phi (x)^\vartheta \rangle
\end{equation}
\begin{equation}
|\phi (y)^{\vartheta +1}\rangle=\left(
\begin{array}{cc}
\cos \triangle \sigma^\vartheta  & -\sin \triangle \sigma^\vartheta  \\
\sin \triangle \sigma^\vartheta  & \cos \triangle \sigma^\vartheta \\
\end{array}
\right)|\phi (y)^\vartheta \rangle
\end{equation}
 where,
 \begin{equation}
 \rho^{\vartheta +1}=\rho^\vartheta  + \triangle \rho^\vartheta 
 \label{eq:alpha}
 \end{equation}
 and
 \begin{equation}
 \sigma^{\vartheta +1}=\sigma^\vartheta  + \triangle \sigma^\vartheta 
 \label{eq:beta}
 \end{equation}
Here, Equations~\ref{eq:alpha} and~\ref{eq:beta} measure the phase change or angles $\triangle \rho^\vartheta $ and $\triangle \sigma^\vartheta $ for the dressed quantum layer in DSQ-Net at epoch $\vartheta $, respectively. Consider 
\begin{equation}
\mathcal{A}^\vartheta =\rho^\vartheta - \overline {\rho^\vartheta }
\end{equation}
\begin{equation}
\mathcal{B}^\vartheta =\sigma^\vartheta - \overline {\sigma^\vartheta }
\end{equation}
and
\begin{equation}
\mathcal{W}^\vartheta =\rho^{\vartheta +1}-\rho^\vartheta =\mathcal{A}^{\vartheta +1}-\mathcal{A}^\vartheta 
\end{equation}
\begin{equation}
\mathcal{V}^\vartheta =\sigma^{\vartheta +1}-\sigma^\vartheta =\mathcal{B}^{\vartheta +1}-\mathcal{B}^\vartheta 
\end{equation}
Here, the optimal phase or angles for the rotation gates $\mathcal{R}_x$ and $\mathcal{R}_y$ are $\overline \rho^\vartheta $ and $\overline \sigma^\vartheta $, respectively. Concerning $\rho_j^\vartheta$ and $\sigma_j^\vartheta$, the loss function $\mathcal{L} (\rho,\sigma)$is differentiated as follows.
 \begin{equation}
 \frac{\partial \mathcal{L} (\rho,\sigma)}{\partial \rho_j^\vartheta } = \sum_{j=1}^\mathcal{C} \frac{\partial z(\theta_j^\vartheta )}{\rho_j^\vartheta } \left[\frac{p_j}{z(\theta_j^\vartheta )} - \frac{p_j-1}{1-z(\theta_j^\vartheta )} \right]
 \end{equation}
\begin{equation}
 \frac{\partial \mathcal{L} (\rho,\sigma)}{\partial \sigma_j^\vartheta } = \sum_{j=1}^\mathcal{C} \frac{\partial z(\theta_j^\vartheta )}{\sigma_j^\vartheta } \left[\frac{p_j}{z(\theta_j^\vartheta )} - \frac{p_j-1}{1-z(\theta_j^\vartheta )} \right]
 \end{equation}
The gradients of the VQC parameters $\rho$ and $\sigma$ are computed using parameter shift methods, as shown in \cite{mitarai2018}.
\begin{equation}
\frac{\partial z(\theta_j^\vartheta )}{\rho_j^\vartheta } = \frac{1}{2}\left[\mathcal{M}^{\vartheta +1}_{\rho\pm \frac{\pi}{2}}(\phi_j(x))-\mathcal{M}^\vartheta _{\rho\pm \frac{\pi}{2}}(\phi_j(x))\right]
 \end{equation}
 and 
\begin{equation}
\frac{\partial z(\theta_j^\vartheta )}{\sigma_j^\vartheta } = \frac{1}{2}\left[\mathcal{M}^{\vartheta +1}_{\sigma\pm \frac{\pi}{2}}(\phi_j(x))-\mathcal{M}^\vartheta _{\sigma\pm \frac{\pi}{2}}(\phi_j(x))\right]
 \end{equation}
Hence, $\mathcal{M}^\vartheta _{\rho\pm \frac{\pi}{2}}(\phi_j(x))$ and $\mathcal{M}^\vartheta _{\sigma\pm \frac{\pi}{2}}(\phi_j(x))$ are measured qubits $\phi_j(x)$ with $\rho_j^\vartheta $ and $\sigma_j^\vartheta $ rotation angles, respectively. The changes in phase or angles for the rotation gate involved in updating the qubits are denoted by $\triangle \rho_j^\vartheta $ and $\triangle \sigma_j^\vartheta $, respectively. Finally, the following formula is used to update rotation angles.
\begin{equation}
\triangle \rho_j^\vartheta =-\eta^\vartheta  \{\frac{\partial z(\theta_j^\vartheta )}{\rho_j^\vartheta }\}
\end{equation}
\begin{equation}
\triangle \sigma_j^\vartheta =-\gamma^\vartheta  \{\frac{\partial z(\theta_j^\vartheta )}{\sigma_j^\vartheta }\}
\end{equation}
The learning rates in the gradient descent process for updating the rotation angles are $\eta^\vartheta $ and $\gamma^\vartheta $.

\footnotesize
\subsection{Data availability}
\label{data}
 The MNIST\cite{lecun1998}, FashionMNIST\cite{xiao2017} and KMNIST\cite{clanuwat2018} data sets can be found in the following links:\url{https://github.com/zalandoresearch/FashionMNIST},  \url{http://yann.lecun.com/exdb/mnist/} and \url{https://github.com/rois-codh/kmnist}, respectively.
 
\subsection{Code Availability}
\label{code}
The PyTorch source code \url{https://github.com/darthsimpus/RQNN} for executing the RQNN is made available on GitHub to reproduce the results reported in the manuscript.

\section*{Acknowledgements}
This work was partially supported by the Center of Advanced Systems Understanding (CASUS), financed by Germany's Federal Ministry of Education and Research (BMBF) and by the Saxon state government out of the state budget approved by the Saxon State Parliament.

\section*{Author contributions statement}
D.~Konar, S.~Bhandary and A.~D.~Sarma conceived the main methodology, implementations, experiments and analyzed the results, and prepared the original draft. E.~Gelenbe developed the RNN, provided the theoretical framework and edited the manuscript. A.~ Cangi analyzed the results and edited the manuscript. 

\section*{Competing interests} 
The authors hereby declare that no conflict of financial/personal interest or belief could affect their objectivity.
\end{document}